\documentclass[12pt]{article}
\usepackage{amssymb,amsmath,epsfig}

\renewcommand{\theequation}{\arabic{section}.\arabic{equation}}

\begin{document}

\title{\bf Isothermal Plasma Wave Properties of the Schwarzschild de-Sitter Black Hole
in a Veselago Medium}
\author{M. Sharif \thanks{msharif@math.pu.edu.pk} and Ifra Noureen
\thanks{ifra.noureen@gmail.com}\\
Department of Mathematics, University of the Punjab,\\
Quaid-e-Azam Campus, Lahore-54590, Pakistan.}
\date{}
\maketitle
\begin{abstract}
In this paper, we study wave properties of isothermal plasma for
the Schwarzschild de-Sitter black hole in a Veselago medium. We
use ADM $3+1$ formalism to formulate general relativistic
magnetohydrodynamical (GRMHD) equations for the Schwarzschild
de-Sitter spacetime in Rindler coordinates. Further, Fourier
analysis of the linearly perturbed GRMHD equations for the
rotating (non-magnetized and magnetized) background is taken whose
determinant leads to a dispersion relation. We investigate wave
properties by using graphical representation of the wave vector,
the refractive index, change in refractive index, phase and group
velocities. Also, the modes of wave dispersion are explored. The
results indicate the existence of the Veselago medium.
\end{abstract}

\section{Introduction}

Our solar system is filled with a wide range of celestial objects.
Black hole is one of such objects, having so strong gravitational
pull that no nearby matter or radiation, not even light can escape
from its gravitational field. Astronomers are curious to extract
real life examples of black hole. The presence of matter in the
form of white dwarfs and neutron stars suggests the existence of
stellar mass black holes. The accumulated evidence for the black
hole existence is now very captivating. It is believed that
collapse of a massive star under its own gravity leads to the
formation of black hole (Das 2004). Plasmas are abundant in
nature, almost found everywhere in an interstellar medium. It is a
distinct state of matter with free electric charge carriers which
behave collectively and respond strongly to electromagnetic fields
(Raine and Thomas 2005). Black hole (in its surroundings) attracts
plasma towards the event horizon due to its strong gravitational
pull. The plasma forms an accretion disk due to interaction of
plasma field with black hole gravity.

The theory of general relativistic magnetohydrodynamics (GRMHD) is
the most reliable discipline to examine the dynamics of magnetized
plasma and effects of black hole gravity. The de-Sitter spacetime
is a vacuum solution of the Einstein field equations including a
positive cosmological constant (Rindler 2001). The Schwarzschild
de-Sitter (SdS) metric describes a black hole expressing a patch
of the de-Sitter spacetime. Since the SdS black hole is
non-rotating, so plasma in magnetosphere moves only along the
radial direction. According to the recent cosmological and
astrophysical observations, our universe is accelerating rather
than decelerating and inclusion of positive cosmological constant
reveals the expanding universe (Reiss et al.1998; Bahcall et al.
1999; Perlmutter et al. 1997). That is why our universe approaches
to de-Sitter universe in future. This motivates the study of
plasma waves in de-Sitter spacetime.

Petterson (1974) investigated the strong gravitational field close
to the surface of compact objects for the Schwarzschild black
hole. Narayan (2005) suggested that compact objects having mass
three times the solar mass can be identified as black hole
candidates. Plasma present in magnetosphere is perturbed by
gravity of black hole. Zerilli (1970a, 1970b, 1970c) used linear
perturbation to explore gravitational field of a particle falling
in the Schwarzschild black hole. Price (1972a, 1972b) discussed
dynamics of approximately spherical star by using non-spherical
perturbations. Regge and Wheeler (1957) also used non-spherical
perturbation to investigate the stability of Schwarzschild
singularity. Gleiser et al. (1972) explored the stability of black
holes by considering second order perturbations.

Arnowitt, Deser and Misner (ADM) (1962) proposed $3+1$ formalism
for an easy approach to General Relativity (GR) by separating
metric field into two parts (space and time) to characterize the
coordinate system. Smarr and York (1978) used this formulation to
explore spacetime kinematics numerically. Israel (1967, 1968)
discussed event horizons in static vacuum and static
electro-vacuum spacetimes. Thorne and Macdonald (1982a, 1982b)
explained how $3+1$ split is appropriate approach for black hole
theory. Macdonald and Suen (1985) developed a self-consistent
formalism to treat electromagnetic and gravitational fields near
black hole horizon. Sakai and Kawata (1980) analyzed wave
propagation in ultra-relativistic plasma, parallel to a constant
magnetic field in a frame of two fluid model. Holcomb (1990) and
Dettmann et al. (1993) constructed electrodynamical equations for
the universe models. Holcomb and Tajima (1989) formulated
linearized theory for relativistic plasma and found results for
matter fluctuations in the early universe.

Rezolla et al. (2003) explored dynamics of thick disks around SdS
black hole by considering the effects of cosmological constant.
Font and Daigne (2002) studied stability of thick accretion disks
around black holes. Myung (2001) developed entropy bounds for SdS
black hole. Suneeta (2003) considered quasinormal modes for scalar
field perturbations of SdS black hole. Setare (2005) obtained area
and entropy spectrum near extremal SdS black hole horizon. Zhang
(1989a) modified the stationary symmetric GRMHD black hole
configuration theory. The same author (Zhang 1989b) explored the
modes of perturbation in rotating black hole. Buzzi et al. (1995a,
1995b) determined the properties of waves propagating in two fluid
plasma for the Schwarzschild black hole. Ali and Rahman (2009)
explained transverse wave propagation in two fluid plasma around
SdS black hole. Sharif and his collaborators (Sharif and Sheikh
2007a, 2007b, 2007c, 2008a, 2008b, 2008c 2009a, 2009b; Sharif and
Mustafa 2008; Sharif and Rafique 2010) have explored wave
properties of cold, isothermal and hot plasmas with Schwarzschild
as well as Kerr spacetimes in the usual medium.

The medium with both negative permeability and permittivity has
the unusual electromagnetic properties named as Veselago medium or
negative index medium (NIM), after a Russian physicist Veselago
(1968). It is also called as double negative medium (DNM) or
negative phase velocity medium (NPV). Valanju et al. (2002)
presented treatment for refraction of electromagnetic waves in a
NIM. Ross et al. (2006) concluded that propensity of a rotating
black hole is enhanced in the presence of charge to support wave
propagation with negative phase velocity in its ergosphere.
Ziolkowski and Heyman (2001) studied wave propagation analytically
and numerically in NIM. Nagar et al. (2004) reported results from
numerical simulations of gravitational radiations emitted due to
matter accretion from non-rotating compact objects. In recent
papers, Sharif and Mukhtar (2011a, 2011b) have discussed wave
properties with non-rotating as well as rotating background
plasmas (isothermal and hot) in this unusual medium.

This paper deals with wave properties of isothermal plasma around
SdS black hole in a Veselago medium. We consider $3+1$ GRMHD
equations and determine a dispersion relation by Fourier analysis
for both magnetized and non-magnetized backgrounds. The results are
discussed by three dimensional plot of wave vector, refractive index
and change in refractive index. The paper is organized as follows:
In section \textbf{2}, linearly perturbed $3+1$ GRMHD equations for
isothermal plasma and their Fourier analysis is developed. Sections
\textbf{3} and \textbf{4} provide reduced form of the GRMHD
equations for rotating (non-magnetized and magnetized respectively)
plasmas. We summarize our results in the last section.

\section{GRMHD Equations in a Veselogo Medium With Isothermal Plasma Assumption}

The general line element in ADM $3+1$ formalism is given as
follows (Zhang 1989b)
\begin{equation}\setcounter{equation}{1}\label{1}
ds^2=-\alpha^2dt^2+\eta_{ij}(dx^i+\beta^idt)(dx^j+\beta^jdt).
\end{equation}
A natural observer associated with this spacetime is known as
fiducial observer (FIDO), $\alpha$ denotes lapse function (ratio
of FIDO proper time to universal time i.e., $\frac{d\tau}{dt}$),
$\beta^i$ is three-dimensional shift vector (which determines
change in spatial coordinates) and $\eta_{ij}~(i,j=1,2,3)$ are the
components of three-dimensional hypersurfaces. The SdS spacetime
in Rindler coordinates is given by (Ali and Rehman 2009)
\begin{equation}\label{2}
ds^2=-\alpha^2(z)dt^2+dx^2+dy^2+dz^2,
\end{equation}
where the directions $z,~y$ and $x$ are analogous to the
Schwarzschild coordinates $r,~\phi$ and $\theta$ respectively. Since
SdS black hole is non-rotating, the shift vector vanishes. On
comparing Eqs.(\ref{1}) and (\ref{2}), we have
\begin{equation}\label{3}
\alpha=\alpha(z),\quad\beta=0,\quad\eta_{ij}=1~(i=j).
\end{equation}

The $3+1$ GRMHD equations for the line element (\ref{2}) in a
Veselago medium are given by Eqs.(\ref{49})-(\ref{53}) in
Appendix. The equation of state for isothermal plasma is (Zhang
1989a)
\begin{eqnarray}\label{3}
\mu=\frac{\rho+p}{\rho_0}=constant,
\end{eqnarray}
here $\rho_0,~\rho,~p$ and $\mu$ denote rest mass density, moving
mass density, pressure and specific enthalpy respectively. The
specific enthalpy is constant while pressure $p\neq0$ for the
isothermal plasma. This equation shows that there is no energy
exchange between plasma and magnetic field of fluid. The
corresponding $3+1$ GRMHD equations ((\ref{49})-(\ref{53})) for
isothermal plasma surrounding the SdS black hole become
\begin{eqnarray}
\label{4} &&\frac{\partial \textbf{B}}{\partial
t}=-\nabla\times(\alpha \textbf{V}\times \textbf{B}),\\
\label{5}&&\nabla.\textbf{B}=0,\\
\label{6} &&\frac{\partial (\rho+p) }{\partial t}+(\rho+p)\gamma^2
\textbf{V}. \frac{\partial \textbf{V}}{\partial t}+(\alpha
\textbf{V}.\nabla)(\rho+p)+(\rho+p)\gamma^2 V.(\alpha
\textbf{V}.\nabla)
\textbf{V}\nonumber\\
&&+(\rho+p) \nabla.(\alpha\textbf{V})=0,\\
\label{7}
&&\left\{\left((\rho+p)\gamma^2+\frac{\textbf{B}^2}{4\pi}\right)\delta_{ij}
+(\rho+p)\gamma^4V_iV_j-\frac{1}{4\pi}B_iB_j\right\}
\left(\frac{1}{\alpha}\frac{\partial}{\partial
t}\right.\nonumber\\
&&\left.+\textbf{V}.\nabla\right)V^j-\left(\frac{\textbf{B}^2}{4\pi}\delta_{ij}-
\frac{1}{4\pi}B_iB_j\right)V^j_{,k}V^k\nonumber\\
&&+(\rho+p)\gamma^2a_i+p_{,i}=\frac{1}{4\pi}
(\textbf{V}\times\textbf{B})_i\nabla.(\textbf{V}\times\textbf{B})
-\frac{1}{8\pi\alpha^2}(\alpha\textbf{B})^2_{,i}\nonumber\\
&&+\frac{1}{4\pi\alpha}(\alpha B_i)_{,j}B^j-\frac{1}{4\pi\alpha}
[\textbf{B}\times\{\textbf{V}\times(\nabla\times(\alpha\textbf{V}\times\textbf{B}))\}]_i,\\
\label{8} &&(\frac{1}{\alpha}\frac{\partial}{\partial
t}+\textbf{V}.\nabla)(\rho+p)\gamma^2-\frac{1}{\alpha}\frac
{\partial p}{\partial t}+2(\rho+p)\gamma^2(\textbf{V}.\textbf{a})
+(\rho+p)\nonumber\\
&&\gamma^2(\nabla.\textbf{V})
-\frac{1}{4\pi\alpha}\left.(\textbf{V}\times\textbf{B}).(\textbf{V}\times\frac
{\partial \textbf{B}}{\partial t}\right.)
-\frac{1}{4\pi\alpha}\left.(\textbf{V}\times\textbf{B}).(\textbf{B}\times\frac{\partial
\textbf{V}}{\partial
t}\right.)\nonumber\\&&+\frac{1}{4\pi\alpha}\left(\textbf{V}\times\textbf{B}).
(\nabla\times\alpha\textbf{B}\right.)=0.
\end{eqnarray}
In rotating background, we assume that plasma flow is in two
dimensions, i.e., in $xz$-plane. Therefore FIDO's measured
velocity $\textbf{V}$ and magnetic field $\textbf{B}$ turn out to
be
\begin{eqnarray}\label{9}
\textbf{V}=V(z)\textbf{e}_x+u(z)\textbf{e}_z,\quad
\textbf{B}=B[\lambda(z)\textbf{e}_x+\textbf{e}_z],
\end{eqnarray}
here $B$ is an arbitrary constant. The relation between the
quantities $\lambda,~u$ and $V$ is given by (Sharif and Sheikh
2007a, 2007b, 2008a, 2008b, 2008c)
\begin{equation}\label{a}
V=\frac{V^F}{\alpha}+\lambda u,
\end{equation}
where $V^F$ is an integration constant. The Lorentz factor,
$\gamma=\frac{1}{\sqrt{1-\textbf{V}^2}}$ becomes
\begin{equation}\label{b}
\gamma=\frac{1}{\sqrt{1-u^2-V^2}}.
\end{equation}

When the plasma flow is perturbed due to black hole gravity, we
use linear perturbation. The flow variables (mass density $\rho$,
pressure $p$, velocity $\textbf{V}$ and magnetic field
$\textbf{B}$) take the form
\begin{eqnarray}\label{10}
&&\rho=\rho^0+\delta\rho=\rho^0+\rho\widetilde{\rho},\quad
p=p^0+\delta p=p^0+p\widetilde{p},\nonumber\\
&&\textbf{V}=\textbf{V}^0+\delta\textbf{V}=\textbf{V}^0+\textbf{v},~
\textbf{B}=\textbf{B}^0+\delta\textbf{B}=\textbf{B}^0+B\textbf{b},
\end{eqnarray}
where unperturbed quantities are denoted by
$\rho^0,~p,~\textbf{V}^0$ and $~\textbf{B}^0$, the linearly
perturbed quantities are represented by $\delta\rho,~\delta
p,~\delta\textbf{V}$ and $\delta\textbf{B}$. We introduce the
following dimensionless quantities
$\widetilde{\rho},~\widetilde{p},~v_x,~v_z,~b_x$ and $b_z$ for the
perturbed quantities
\begin{eqnarray}\label{11}
&&\tilde{\rho}=\tilde{\rho}(t,z),\quad
\tilde{p}=\tilde{p}(t,z),\quad\textbf{v}=\delta\textbf{V}=v_x(t,z)\textbf{e}_x
+v_z(t,z)\textbf{e}_z,\nonumber\\
&&\textbf{b}=\frac{\delta\textbf{B}}{B}=b_x(t,z)\textbf{e}_x
+b_z(t,z)\textbf{e}_z.
\end{eqnarray}
When we insert these linear perturbations in the perfect GRMHD
equations (Eqs.(\ref{4})-(\ref{8})) along with Eq.(\ref{11}), the
component form of these equations will be (Sharif and Mukthar
2011a, 2011b)
\begin{eqnarray}
\label{17}&&\frac{1}{\alpha}\frac{\partial b_x}{\partial
t}-ub_{x,z}=(ub_x-Vb_z-v_x+\lambda v_z)\nabla
\ln\alpha\nonumber\\
&&-(v_{x,z}-\lambda
v_{z,z}-\lambda'v_z+V'b_z+Vb_{z,z}-u'b_x),\\\label{18}
&&\frac{1}{\alpha}\frac{\partial b_z}{\partial t}=0,\\\label{19}
&&b_{z,z}=0,
\end{eqnarray}
\begin{eqnarray}
\label{20} &&\rho\frac{\partial\tilde{\rho}}{\partial
t}+p\frac{\partial\tilde{p}}{\partial
t}+(\rho+p)\gamma^2(V\frac{\partial v_x}{\partial
t}+u\frac{\partial v_z}{\partial t})+\alpha u\rho\rho_{,z}+\alpha
upp_{,z}\nonumber\\
&&+\alpha(\rho+p)\{\gamma^2uVv_{x,z}+(1+\gamma^2u^2)v_{z,z}\}-\frac{1}{\gamma}
(\tilde{\rho}-\tilde{p})(\alpha u\gamma p)_{,z}\nonumber\\
&&+\alpha(\rho+p)\gamma^2u
\{(1+2\gamma^2V^2)V'+2\gamma^2uVu'\}v_x-\alpha(\rho+p)\nonumber\\
&&\times\{(1-2\gamma^2u^2)(1+\gamma^2u^2)\frac{u'}{u}\}
-2\gamma^4u^2VV'\}v_z=0,\\\label{21}
&&\left\{(\rho+p)\gamma^2(1+\gamma^2V^2)
+\frac{B^2}{4\pi}\right\}\frac{1}{\alpha}\frac{\partial
v_x}{\partial t}+\left\{(\rho+p)\gamma^4uV-\frac{\lambda B
^2}{4\pi}\right\}\nonumber\\
&&\times\frac{1}{\alpha}\frac{\partial v_z}{\partial
t}+\left\{(\rho+p)\gamma^2(1+\gamma^2V^2)
+\frac{B^2}{4\pi}\right\}uv_{x,z}+\left\{(\rho+p)\gamma^4uV\right.\nonumber\\
&&\left.-\frac{\lambda B^2}{4\pi}\right\}uv_{z,z}
-\frac{B^2}{4\pi}(1+u^2)b_{x,z}-\frac{B^2}{4\pi\alpha}\left\{\alpha'(1+u^2)+\alpha
uu'\right\}b_x\nonumber\\
&&+\gamma^2u(\rho\tilde{\rho}+p\tilde{p})\left\{(1+\gamma^2V^2)V'+\gamma^2uVu'\right\}
+[(\rho+p)\gamma^4u\{(1\nonumber\\
&&+4\gamma^2V^2)uu'+4VV'(1+\gamma^2V^2)\}+\frac{B^2u\alpha'}{4\pi\alpha}]v_x+[(\rho+p)
\gamma^2\{(1\nonumber\\
&&+2\gamma^2u^2)(1+2\gamma^2V^2)V'-\gamma^2V^2V'
+2\gamma^2(1+2\gamma^2u^2)uVu'\}\nonumber\\
&&-\frac{B^2u} {4\pi\alpha}(\lambda\alpha)']v_z=0,\\\label{22}
&&\left\{(\rho+p)\gamma^2(1+\gamma^2u^2)
+\frac{\lambda^2B^2}{4\pi}\right\}\frac{1}{\alpha}\frac{\partial
v_z}{\partial t}+\left\{(\rho+p)\gamma^4uV -\frac{\lambda B
^2}{4\pi}\right\}\nonumber\\
&&\times\frac{1}{\alpha}\frac{\partial v_x}{\partial t}
+\left\{(\rho+p)\gamma^2(1+\gamma^2u^2)+\frac{\lambda^2B^2}{4\pi}\right\}
uv_{z,z}+\{(\rho+p)\gamma^4u\nonumber\\
&&\times V-\frac{\lambda B^2}{4\pi}\}uv_{x,z}+\frac{\lambda
B^2}{4\pi}(1+u^2)b_{x,z}+\frac{B^2}{4\pi\alpha}\{\alpha'\lambda-(\alpha\lambda)'
+u\lambda\nonumber\\
&&\times(u\alpha'+u'\alpha)\}b_x+(\rho\tilde{\rho}+p\tilde{p})\gamma^2\{a_z
+u u'(1+\gamma^2u^2)+\gamma^2u^2VV'\}\nonumber\\
&&+[(\rho+p)\gamma^4\{u^2V'(1+4\gamma^2V^2)+2V(a_z+uu'(1+2\gamma^2u^2))\}-\lambda
B^2\nonumber\\&&\times\frac{u\alpha'}{4\pi\alpha}]v_x+[(\rho+p)
\gamma^2\{u'(1+\gamma^2u^2)(1+4\gamma^2u^2)+2u\gamma^2(a_z+(1\nonumber\\
&&+2\gamma^2u^2)V V')\}+\frac{\lambda
B^2u}{4\pi\alpha}(\alpha\lambda)']v_z+(p'\tilde{p}+p\tilde{p}')=0,
\end{eqnarray}
\begin{eqnarray}
\label{23} &&\frac{1}{\alpha}\gamma^2\rho\frac{\partial
\tilde{\rho}}{\partial t}+\frac{1}{\alpha}\gamma^2p\frac{\partial
\tilde{p}}{\partial
t}+\gamma^2(\rho'+p')v_z+u\gamma^2(\rho\tilde{\rho}_{,z}+p\tilde{p}_{,z}+\rho'\tilde{\rho}\nonumber\\
&&+p'\tilde{p})-\frac{1}{\alpha}p\frac{\partial
\tilde{p}}{\partial
t}+2\gamma^2u(\rho\tilde{\rho}+p\tilde{p})a_z+\gamma^2u'(\rho\tilde{\rho}+p\tilde{p})+2(\rho\nonumber\\
&&+p)\gamma^4(uV'+2uVa_z+u'V)v_x+2(\rho+p)\gamma^2(2\gamma^2uu'+a_z\gamma^4\nonumber\\
&&+2\gamma^2u^2a_z)v_z+2(\rho+p)\gamma^4uVv_{x,z}+(\rho+p)\gamma^2(1+2\gamma^2u^2)\nonumber
v_{z,z}\\ &&-\frac{B^2}{4\pi\alpha}[(V^2+u^2)\lambda
\frac{\partial b_x}{\partial t}+(V^2+u^2)\frac{\partial
b_z}{\partial t}-V(\lambda V\nonumber+u)\frac{\partial
b_x}{\partial t}\\ &&-u(\lambda V+u)\frac{\partial b_z}{\partial
t}]+\frac{B^2}{4\pi\alpha}[(V-\lambda
u)v_{x,t}+\lambda(u\lambda\nonumber-V)v_{z,t}]\nonumber\\
&&-\frac{B^2} {4\pi}(\lambda\lambda'v_z-\lambda'v_x-\lambda'Vb_z+
\lambda'ub_x-V b_{x,z}+u\lambda b_{x,z})=0.
\end{eqnarray}
The following harmonic spacetime dependence of perturbation is
assumed for the Fourier analysis,
\begin{eqnarray}\label{24}
\widetilde{\rho}(t,z)=c_1e^{-\iota(\omega t-kz)},&\quad&
\widetilde{p}(t,z)=c_2e^{-\iota(\omega t-kz)},\nonumber\\
v_z(t,z)=c_3e^{-\iota(\omega t-kz)},&\quad&
v_x(t,z)=c_4e^{-\iota(\omega t-kz)},\nonumber\\
b_z(t,z)=c_5e^{-\iota(\omega t-kz)},&\quad&
b_x(t,z)=c_6e^{-\iota(\omega t-kz)},
\end{eqnarray}
here $\omega$ and $k$ represent the angular frequency and
$z$-component of the wave vector $(0,0,k)$, respectively. The wave
vector can be used to determine refractive index and the
properties of plasma near the event horizon.

The ratio of speed of light when it travels from one medium to
another is said to be refractive index. Frequency dependence
effects in wave propagation refers to dispersion. This describes
relations between wave properties like wave length, angular
frequency, refractive index etc. (Das 2004). Dispersion is said to
be normal if change in the refractive index with respect to
angular frequency is positive, otherwise anomalous. Using
Eq.(\ref{24}) in Eqs.(\ref{17})-(\ref{23}), we get their Fourier
analyzed form
\begin{eqnarray}
\label{25}&&c_{4}(\alpha'+\iota k\alpha)-c_3\
\left\{(\alpha\lambda)'+\iota k\alpha\lambda\ \right\}-c_5(\alpha
V)'+c_6\{(\alpha
u)'+\iota\omega\nonumber\\
&&+\iota ku\alpha\}=0,
\\\label{26}
&&c_5(\frac{-\iota\omega}{\alpha})=0,\\\label{27} &&c_5\iota k=0,
\end{eqnarray}
\begin{eqnarray}
\label{28} &&c_1\{(-\iota\omega+\iota k\alpha
u)\rho-p\gamma^2\alpha u(VV'+u u')-\alpha'up-\alpha u'p-\alpha
up'\}\nonumber\\
&&+c_2\{(-\iota\omega+\iota k\alpha u)p+\alpha'up+\alpha
u'p+\alpha
up'+p\gamma^2\alpha u(VV'+u u')\}\nonumber\\
&&+c_3(\rho+p)[-\iota\omega\gamma^2u+\iota
k\alpha(1+\gamma^2u^2)-\alpha\{(1-2\gamma^2u^2)(1+\gamma^2u^2)
\nonumber\\
&&\times\frac{u'}{u}-2\gamma^4u^2VV'\}]+c_4(\rho+p)[\gamma^2V(-\iota\omega+\iota
k\alpha
u)+\alpha\gamma^2u\{(1\nonumber\\
&&+2\gamma^2V^2)V'+2\gamma^2uVu'\}]=0,
\\\label{29}
&&c_1\rho\gamma^2u\{(1+\gamma^2V^2)V'+\gamma^2uVu'\}+c_2p\gamma^2u\{(1+\gamma^2V^2)V'\nonumber\\
&&+\gamma^2uVu'\}+c_3[-\{(\rho+p)\gamma^4uV-\frac{\lambda\
B^2}{4\pi}\}\frac{\iota\omega}{\alpha}+\{(\rho+p)\gamma^4uV\nonumber\\
&&-\frac{\lambda\ B^2}{4\pi}\}\iota ku+(\rho+p)\gamma^2\{(1+2\gamma^2u^2)(1+2\gamma^2V^2)
-\gamma^2V^2\}V'\nonumber\\
&&+2\gamma^4(\rho+p)uVu'(1+2\gamma^2u^2)-\frac{B^2u}{4\pi\alpha}(\alpha\lambda)']+c_4[-\{(\rho+p)
\gamma^2\nonumber\\
&&(1+\gamma^2V^2)+\frac{B^2}{4\pi}\}\frac{\iota\omega}{\alpha}
+\{(\rho+p)\gamma^2(1+\gamma^2V^2)+\frac{B^2}{4\pi}\}\iota ku\nonumber\\
&&+(\rho+p)\gamma^4u\{(1+4\gamma^2V^2)u
u'+4VV'(1+\gamma^2V^2)\}+\frac{B^2u\alpha'}{4\pi\alpha}]\nonumber\\
&&-c_6\frac{B^2}{4\pi}\{(1+u^2)\iota
k+(1+u^2)\frac{\alpha'}{\alpha}+uu'\}=0,
\\\label{30}
&&c_1\rho\gamma^2\{a_z+uu'(1+\gamma^2u^2)+\gamma^2u^2VV'\}+c_2[p\gamma^2\{a_z+uu'\nonumber\\
&&(1+\gamma^2u^2)+\gamma^2u^2VV'\}+p'+\iota k
p]+c_3[-\{(\rho+p)\gamma^2(1+\gamma^2u^2)\nonumber\\
&&+\frac{\lambda
^2B^2}{4\pi}\}\frac{\iota\omega}{\alpha}+\{(\rho+p)\gamma^2(1+\gamma^2u^2)+\frac{\lambda
^2B^2}{4\pi}\}\iota ku+\{(\rho+p)\gamma^2\nonumber\\
&&\times \{u'(1+\gamma^2u^2)(1+4\gamma^2u^2)+2u\gamma^2\{a_z+(1+2\gamma^2u^2)\}VV'\}\nonumber\\
&&+\frac{\lambda
B^2u}{4\pi\alpha}(\alpha\lambda)']+c_4[-\{(\rho+p)\gamma^4u
V-\frac{\lambda
B^2}{4\pi}\}\frac{\iota\omega}{\alpha}+\{(\rho+p)\gamma^4uV\nonumber\\
&&-\frac{\lambda B^2}{4\pi}\}\iota
ku+\{(\rho+p)\gamma^4\{u^2V'(1+4\gamma^2V^2)+2V
\{(1+2\gamma^2u^2)uu'\nonumber\\
&&+a_z-\frac{\lambda B^2\alpha'
u}{4\pi\alpha}\}]+c_6[\frac{B^2}{4\pi\alpha}\{-(\alpha\lambda)'
+\alpha'\lambda-u\lambda(u\alpha'+u'\alpha)\}\nonumber\\
&&+\frac{\lambda B^2}{4\pi}(1+u^2)\iota k]=0,
\end{eqnarray}
\begin{eqnarray}
\label{31}&&c_1\{(\frac{-\iota\omega}{\alpha}\gamma^2+\iota ku
\gamma^2+2u\gamma^2a_z+\gamma^2u')\rho+u\rho'\gamma^2\}
+c_2\{(\frac{\iota\omega}{\alpha}(1-\gamma^2)\nonumber\\
&&+\iota ku\gamma^2+2\gamma^2ua_z+\gamma^2u')p+u\gamma^2p'\}+c_3\gamma^2\{(\rho'+p')+2\nonumber\\
&&\times(2\gamma^2uu'+a_z+2\gamma^2u^2a_z)(\rho+p)+(1+2\gamma^2u^2)(\rho+p)\iota
k+\frac{\lambda B^2}{4\pi\alpha}\nonumber\\
&&\times(\lambda
u-V)\iota\omega+\alpha\lambda'\}+c_4[2(\rho+p)\gamma^4\{(u
V'+2uVa_z+u'V)+uV\iota
k\}\nonumber\\
&&+\frac{B^2}{4\pi\alpha}(V-u\lambda)\iota\omega-\alpha\lambda']
+c_6[\frac{-B^2}{4\pi\alpha}\{(V^2+u^2)\lambda+
V(\lambda V+u)\iota\omega\}\nonumber\\
&&-\alpha\lambda'u+\iota k\alpha(V-u\lambda)]=0.
\end{eqnarray}

\section{Rotating Non-Magnetized Flow}

For the rotating non-magnetized background of plasma flow, we
substitute $B=0=\lambda$ and $c_5=0=c_6$ in the Fourier analyzed
perturbed GRMHD equations ((\ref{28})-(\ref{31})) (Sharif and
Mukhtar 2011a, 2011b).

\subsection{Numerical Solutions}

For the rotating non-magnetized plasma, we use the following assumptions
to find out the numerical solutions
\begin{enumerate}
\item Specific enthalpy: $\mu=1$,
\item Time lapse: $\alpha=\frac{z}{2r_h}$,
\item  Velocity components: $u=V,~x$ and $z$-components of velocity
yield $u=V=-\frac{1}{\sqrt{z^{2}+2}}$,
\item Stiff fluid: $\rho=p=-\frac{1}{2u}$,
\item Lorentz factor: $ \gamma=\frac{1}{\sqrt{1-u^2-V^2}}=
\frac{\sqrt{z^{2}+2}}{z}$,
\end{enumerate}
where $r_h$ is the SdS event horizon greater than that of the
Schwarzschild event horizon and $r_h\thickapprox
2M\left(1+\frac{4M^{2}}{l^{2}}+...\right)\backsimeq\zeta0.2948km$,
$1\leqslant\zeta\leqslant1.5$ for a black hole mass
$M\thicksim1M_{\bigodot}$. The value of $\zeta$ corresponding to
extremal SdS black hole is $1.5$ (Ali and Rehman 2009).

We consider the region $-5\leq z\leq5$ for wave analysis assuming
that event horizon is at $z=0$. We take this region to explain
waves near horizon only for convenience. Since the flow variables
exhibit large variations in the region $-1\leq z\leq1$, we ignore
it and solve dispersion relation for two meshes, i.e., $-5\leq
z\leq-1$ and $1\leq z\leq5$ (corresponding to near and far
electromagnetic radiation zone). A complex dispersion relation (
Das 2004) is obtained by solving the determinant of the
coefficients of constants of the corresponding equations of the
rotating non-magnetized plasma. The real part of the determinant
yields a quartic equation in $k$
\begin{equation}\label{36}
A_1(z)k^4+A_2(z,\omega)k^3+A_3(z,\omega)k^2+A_4(z,\omega)k+A_5(z,\omega)=0
\end{equation}
which gives four real roots. A cubic equation in $k$ is obtained
from the imaginary part
\begin{equation}\label{37}
B_1(z)k^3+B_2(z,\omega)k^2+B_3(z,\omega)k+B_4(z,\omega)=0
\end{equation}
which yields three real roots. The first and second root of the real
part show wave propagation only in the region $-5\leq z\leq-1$ while
the third and fourth root exhibit waves in the region $1\leq
z\leq5$. The roots of the imaginary part indicate wave propagation
in both meshes $-5\leq z\leq-1$ and $1\leq z\leq5$, i.e., region
near the event horizon and outer end of magnetosphere respectively.

The wave vector, refractive index, its change with respect to
angular frequency, group velocity and phase velocity lead to the
wave properties of the SdS black hole and properties of Veselago
medium. These are shown in Figures \textbf{1-10} by using real
values of $k$ in Eqs.(\ref{36}) and (\ref{37}).

It is given that dispersion is normal if phase velocity is greater
than the group velocity, otherwise anomalous (Achenbach 1973) or
equivalently dispersion is normal if change in refractive index is
positive, anomalous otherwise. We see from figures that some waves
move towards the event horizon and some move away from the
horizon. The dispersion is normal in Figures \textbf{5-7} and
\textbf{10} while it is anomalous in the whole region of Figure
\textbf{8}. The following table classifies the regions of normal
and anomalous dispersion in Figures \textbf{1}-\textbf{4} and
\textbf{9}.
\newpage
\par\noindent
The results deduced from these figures can be expressed in the
following table.

\begin{center}
Table I. Direction and refractive index of waves
\end{center}
\begin{tabular}{|c|c|c|c|c|}
\hline\textbf{Fig.} & \textbf{Direction of Waves} &
\textbf{Refractive Index} ($n$)\\ \hline
& & $n<1$ and decreases in the region \\
\textbf{1} & Move towards the event horizon & $-5\leq z\leq
-1.4,0\leq\omega\leq 2.6$\\&& with the decrease in $z$  \\
\hline
& & $n<1$ and increases in the region\\
\textbf{2} & Move away from the event horizon & $-2.94\leq z\leq
-1.1, 0\leq\omega\leq
3.5$\\&&with the decrease in $z$  \\
\hline
& & $n<1$ and increases in the region\\
\textbf{3} & Move towards the event horizon & $1\leq z\leq1.9,1.6\leq\omega\leq 2.7$\\
& &with the decrease in $z$ \\
\hline
& & $n<1$ and decreases in the region\\
\textbf{4} & Move outwards from the event horizon & $1\leq z\leq5,0\leq\omega\leq 3.8$\\
& &with the decrease in $z$ \\
\hline
& & $n<1$ and decreases in the region\\
\textbf{5} & Move towards the event horizon & $-5\leq z\leq-4,5\leq\omega\leq 8$\\
& &with the decrease in $z$ \\
\hline& & $n<1$ and decreases in the region\\
\textbf{6} & Move away from the event horizon & $1.8\leq z\leq5,5.9\leq\omega\leq 10$\\
& &with the decrease in $z$ \\
\hline& & $n<1$ and decreases in the region\\
\textbf{7} & Move away from the event horizon & $-5\leq z\leq-2.6,0\leq\omega\leq 4.7$\\
& &with the decrease in $z$ \\
\hline& & $n<1$ and decreases in the region\\
\textbf{8} & Move away from the event horizon & $1\leq z\leq5,6.3\leq\omega\leq 10$\\
& &with the decrease in $z$ \\
\hline& & $n<1$ and decreases in the region\\
\textbf{9} & Move towards the event horizon & $-5\leq z\leq-3.4,8.1\leq\omega\leq 10$\\
& &with the decrease in $z$ \\
\hline& & $n<1$ and decreases in the region\\
\textbf{10} & Move toward the event horizon & $1.4\leq z\leq5,0.8\leq\omega\leq 5.4$\\
& &with the decrease in $z$ \\
\hline
\end{tabular}
\begin{figure}
\begin{tabular}{cc}
\epsfig{file=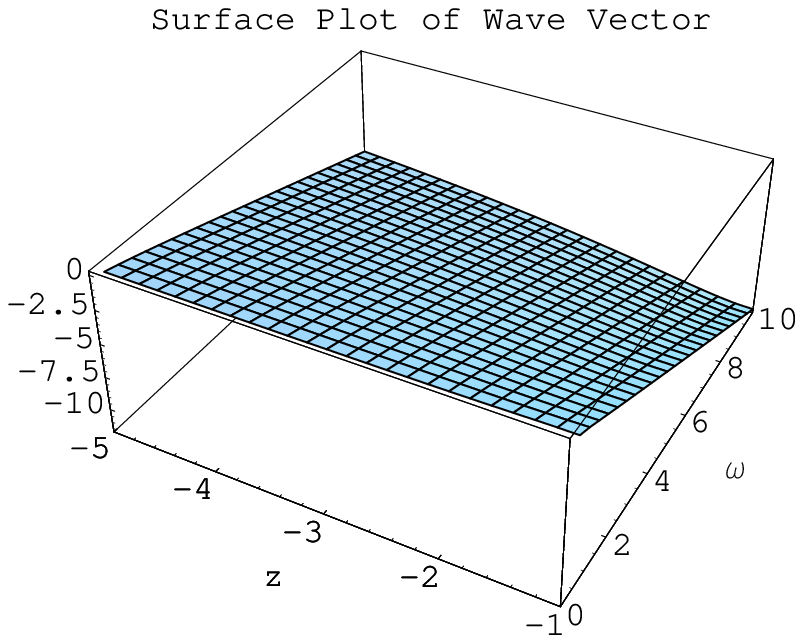,width=0.34\linewidth}
\epsfig{file=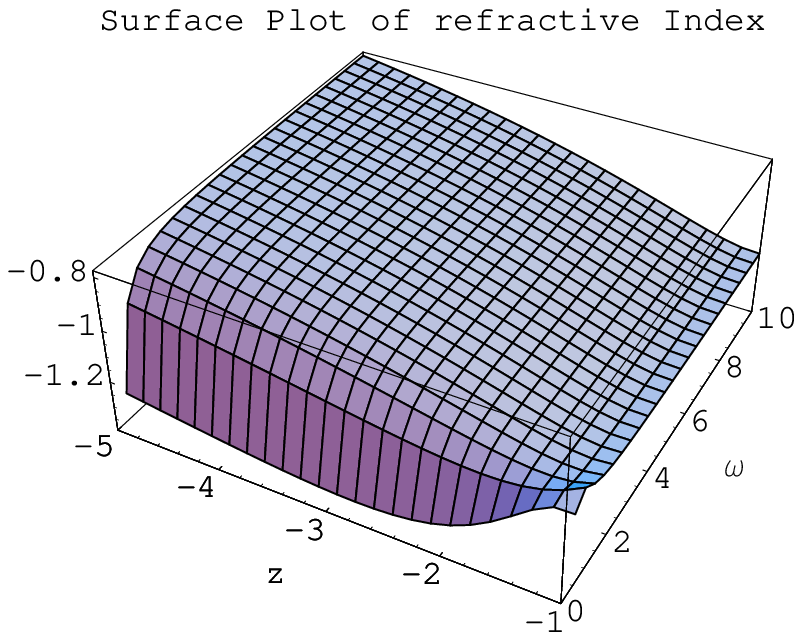,width=0.34\linewidth}\\
\epsfig{file=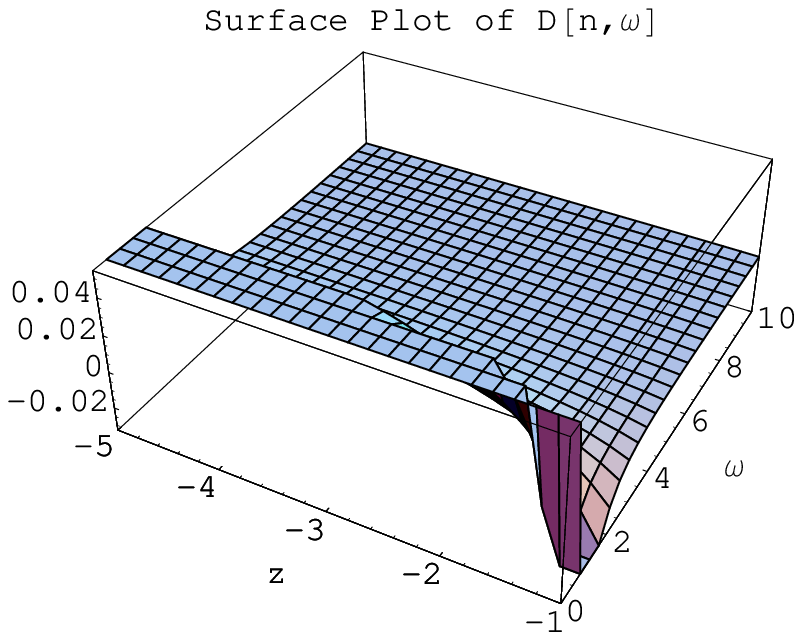,width=0.34\linewidth}
\epsfig{file=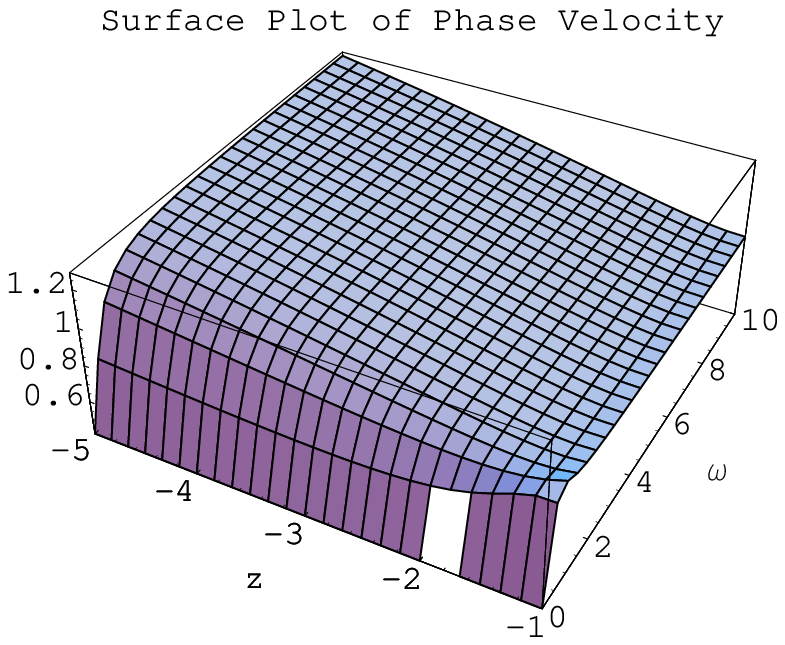,width=0.34\linewidth}
\epsfig{file=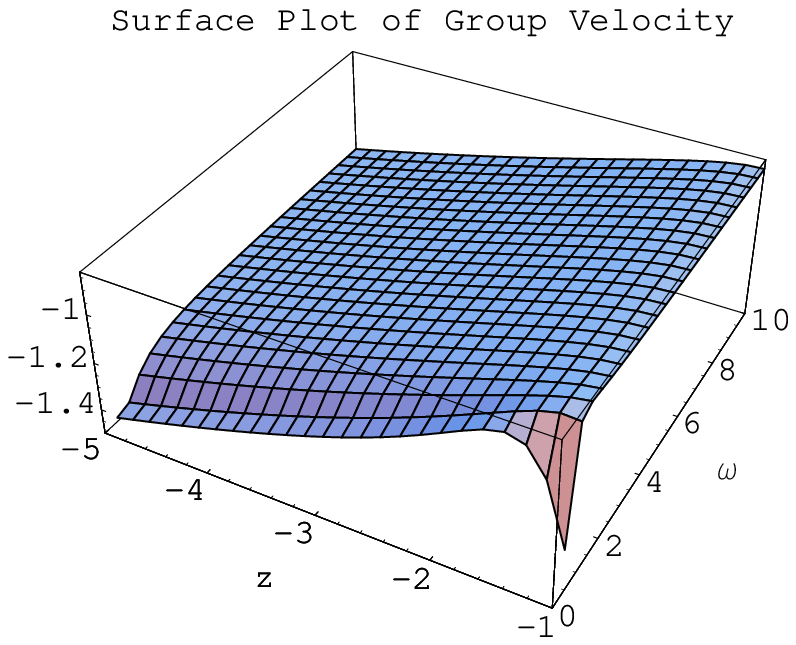,width=0.34\linewidth}\\
\end{tabular}
\caption{Dispersion is normal and anomalous in the region}
\begin{tabular}{cc}\\
\epsfig{file=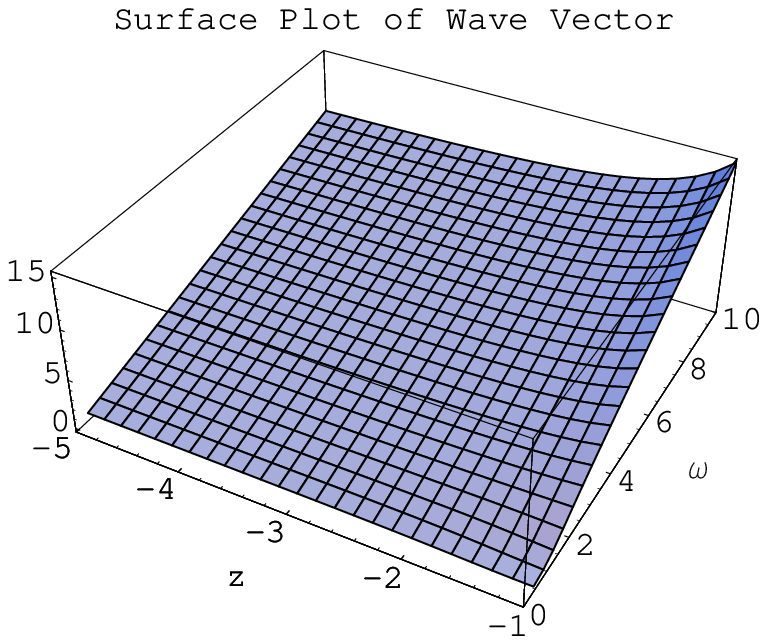,width=0.34\linewidth}
\epsfig{file=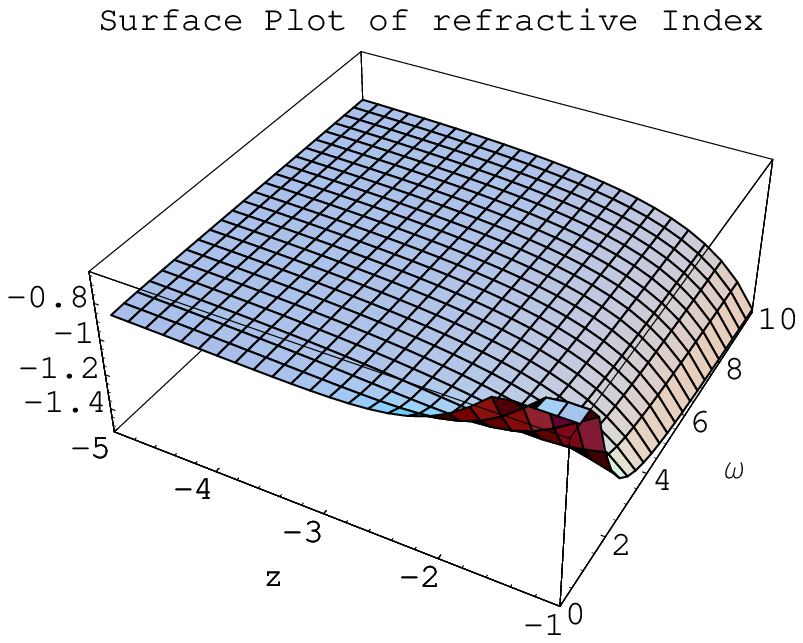,width=0.34\linewidth}\\
\epsfig{file=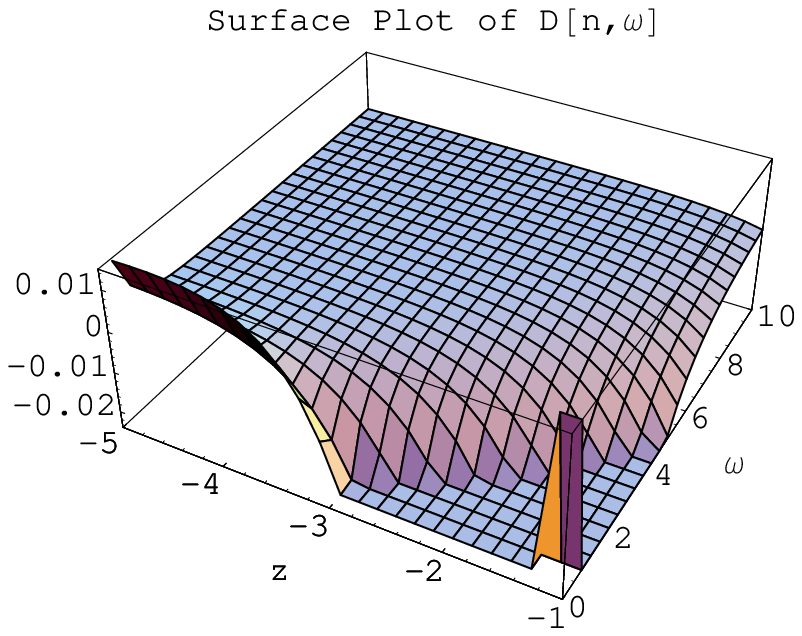,width=0.34\linewidth}
\epsfig{file=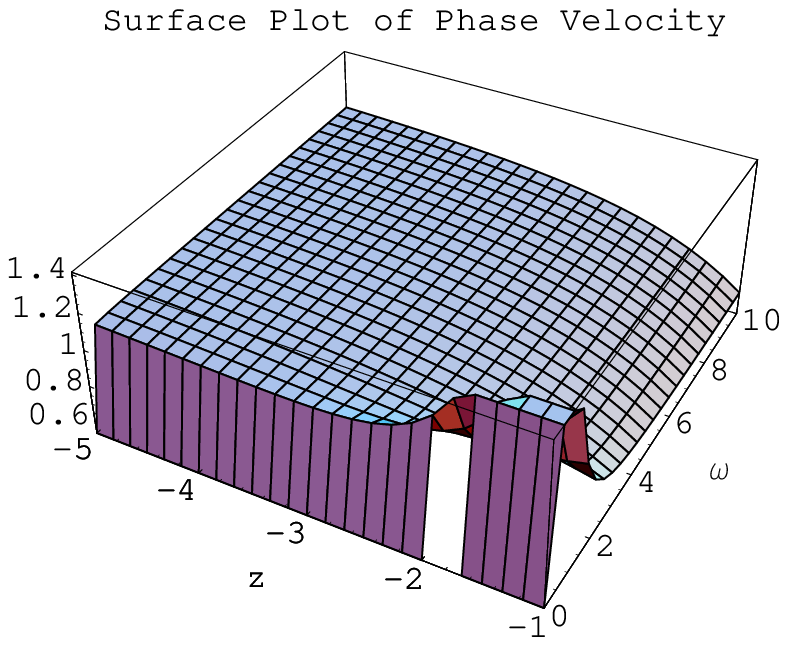,width=0.34\linewidth}
\epsfig{file=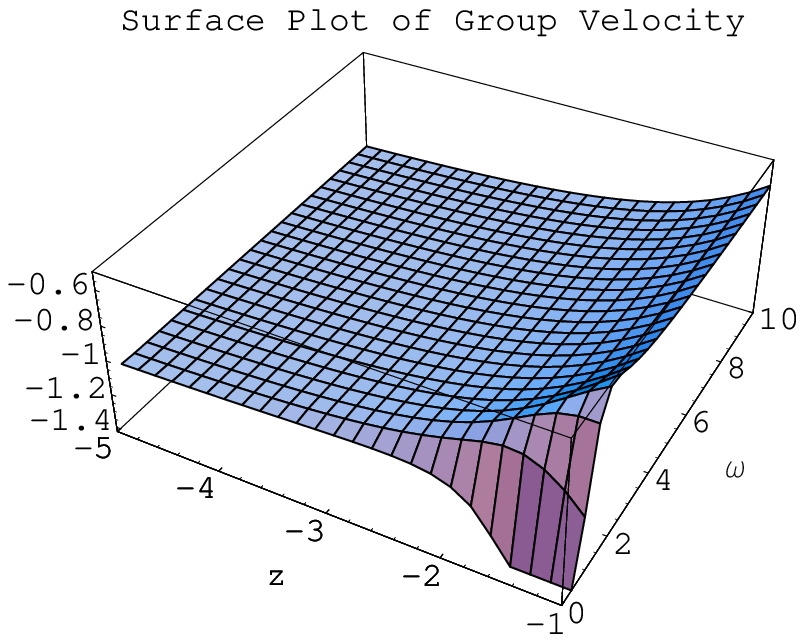,width=0.34\linewidth}\\
\end{tabular}
\caption{Normal as wells as anomalous dispersion occur at random
points in the region.}
\end{figure}
\begin{figure}
\begin{tabular}{cc}
\epsfig{file=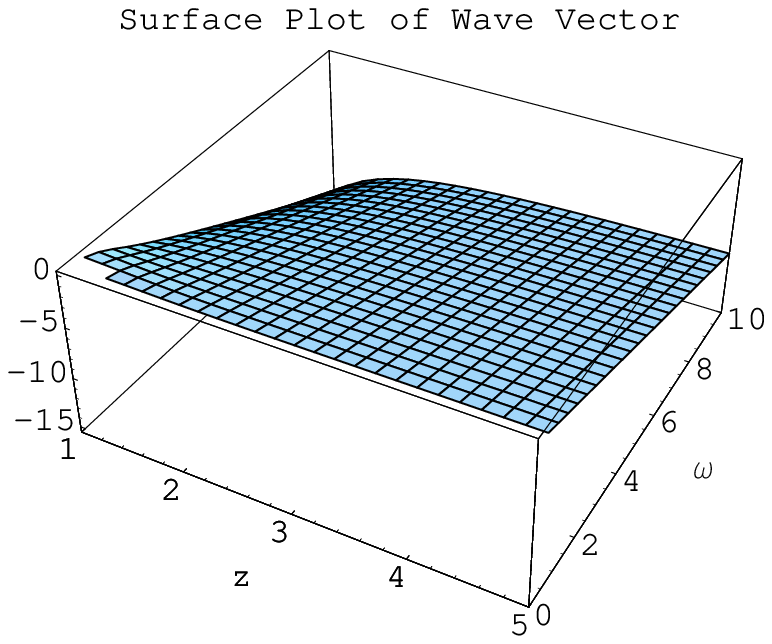,width=0.34\linewidth}
\epsfig{file=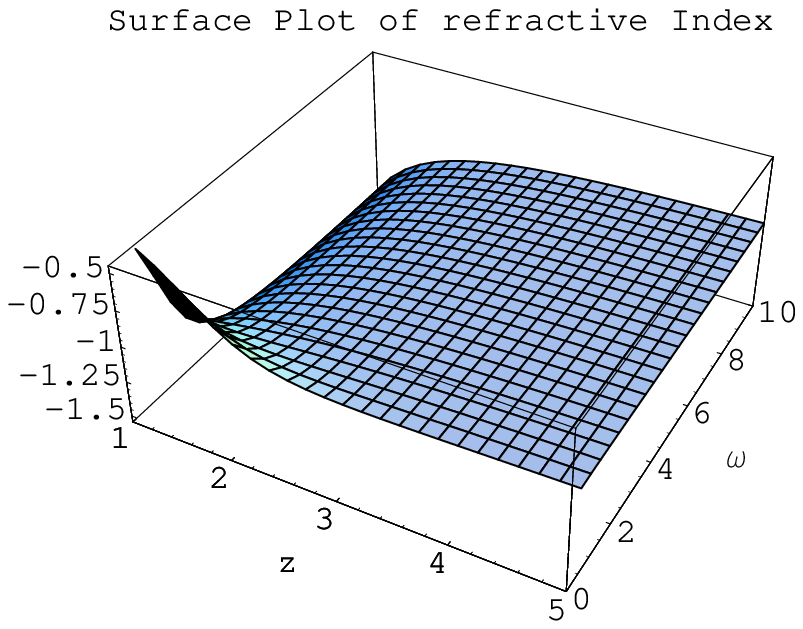,width=0.34\linewidth}\\
\epsfig{file=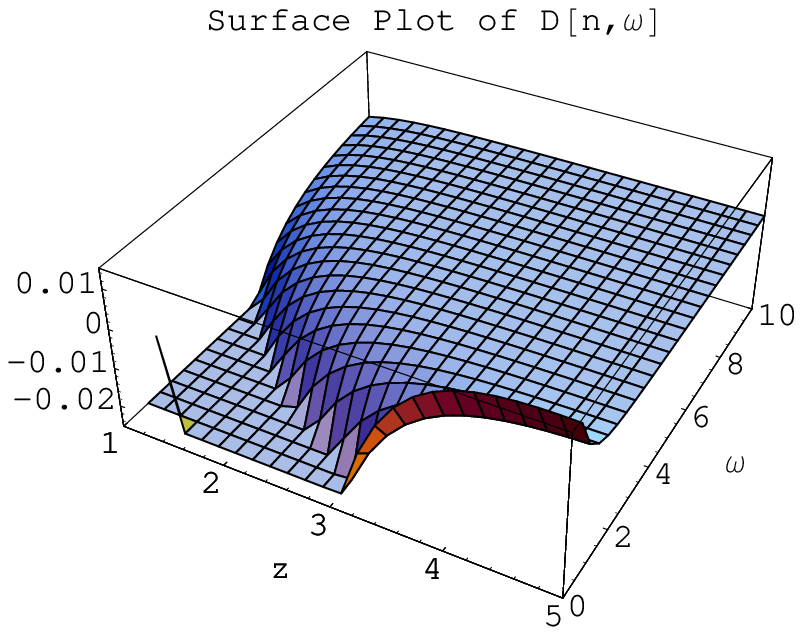,width=0.34\linewidth}
\epsfig{file=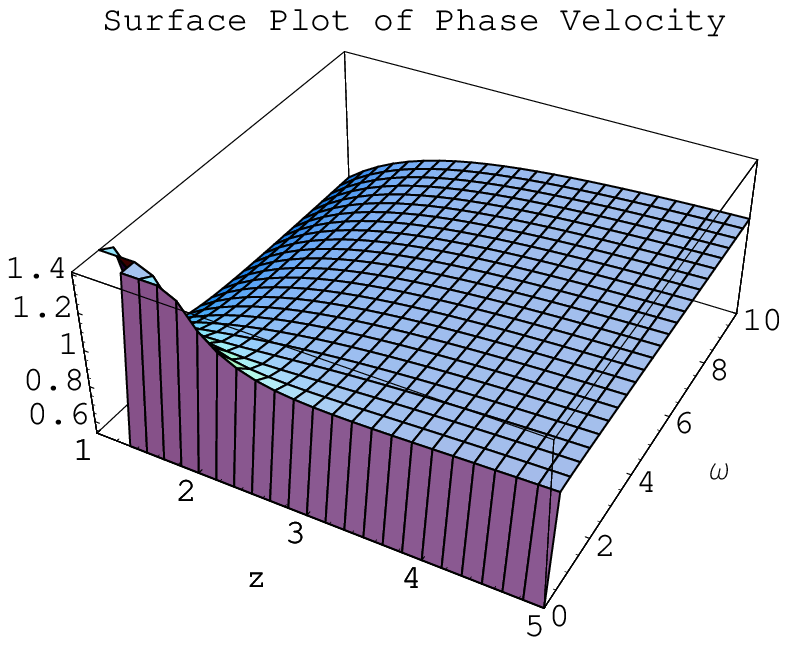,width=0.34\linewidth}
\epsfig{file=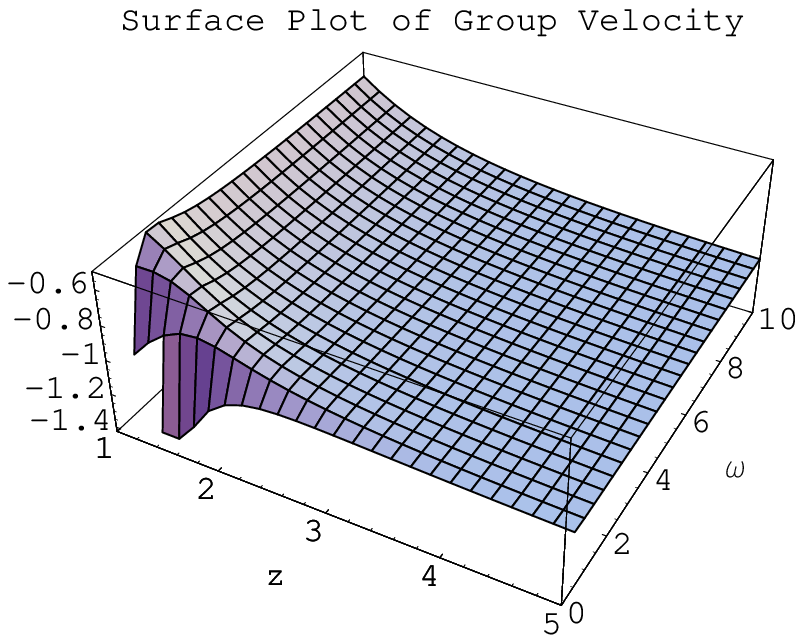,width=0.34\linewidth}\\
\end{tabular}
\caption{Normal and anomalous dispersion of waves is observed.}
\begin{tabular}{cc}\\
\epsfig{file=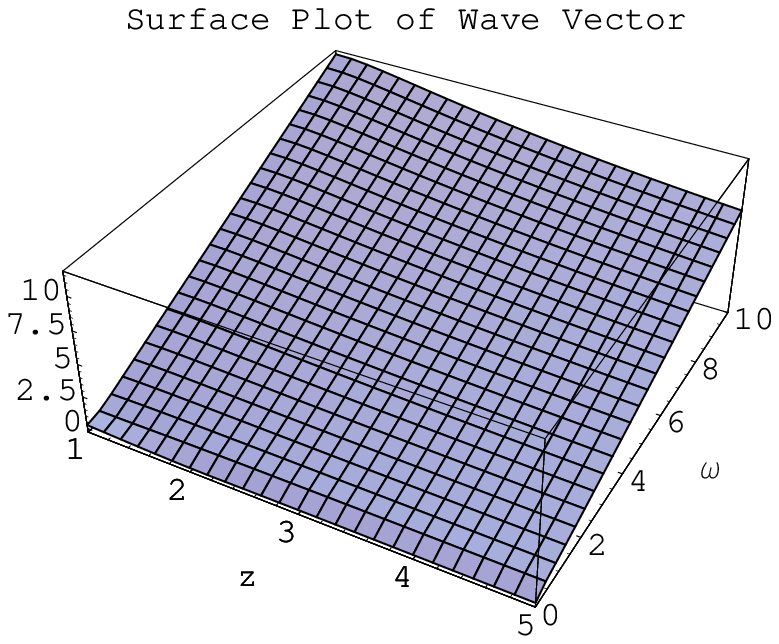,width=0.34\linewidth}
\epsfig{file=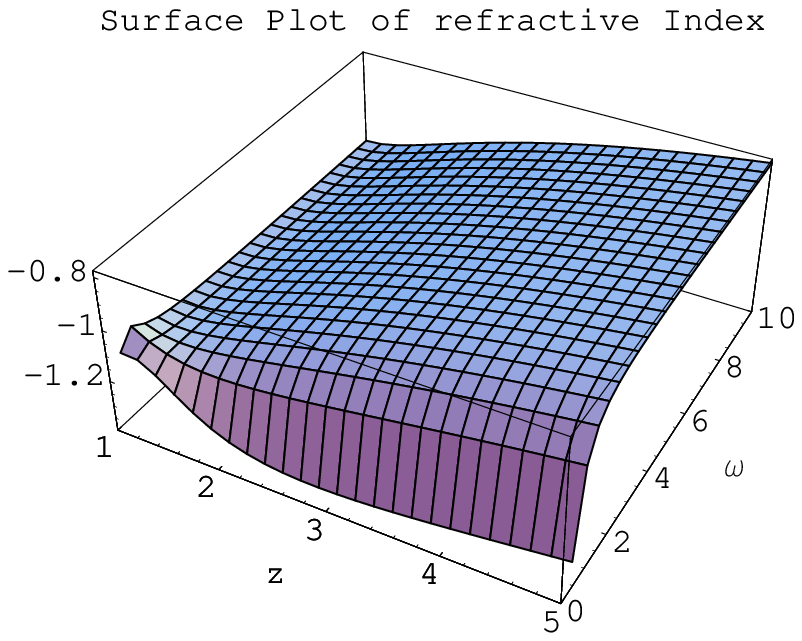,width=0.34\linewidth}\\
\epsfig{file=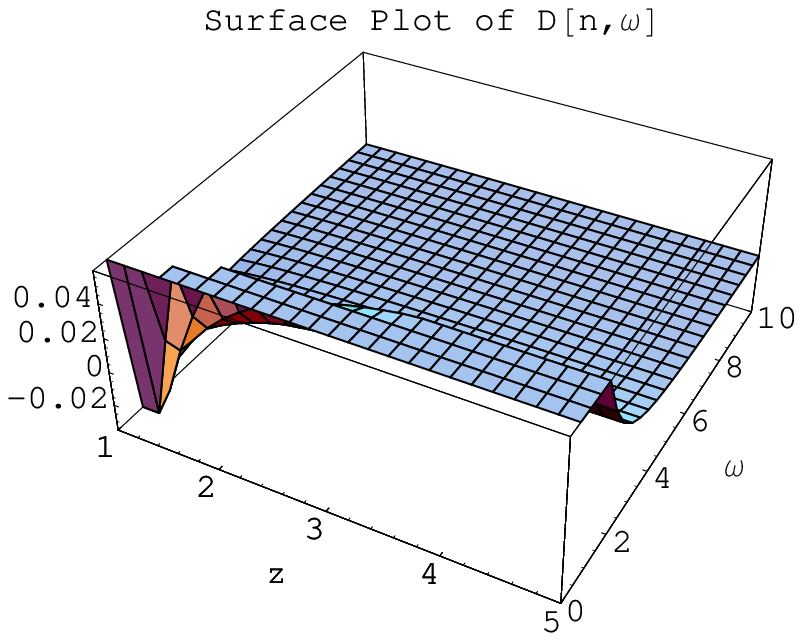,width=0.34\linewidth}
\epsfig{file=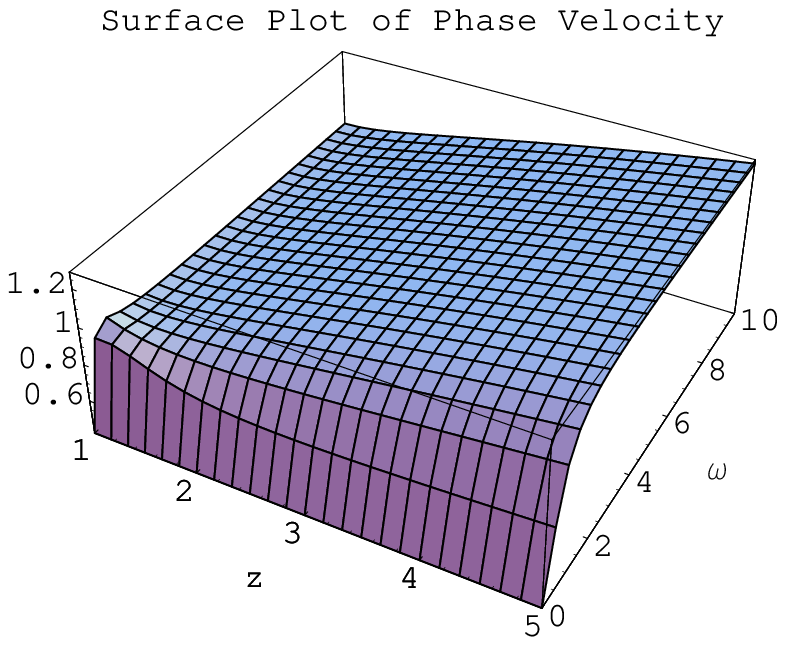,width=0.34\linewidth}
\epsfig{file=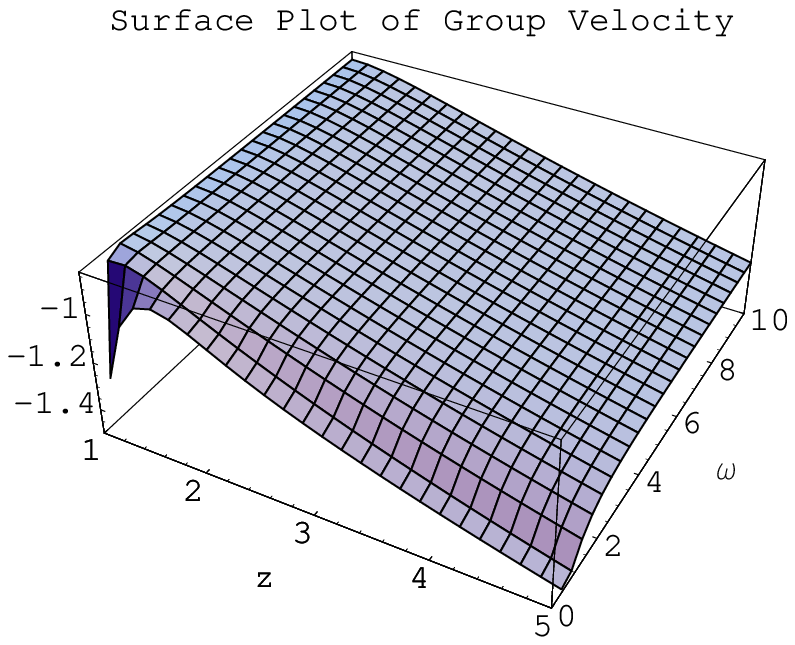,width=0.34\linewidth}\\
\end{tabular}
\caption{Random points of normal and anomalous dispersion are
found in the region.}
\end{figure}
\begin{figure}
\begin{tabular}{cc}
\epsfig{file=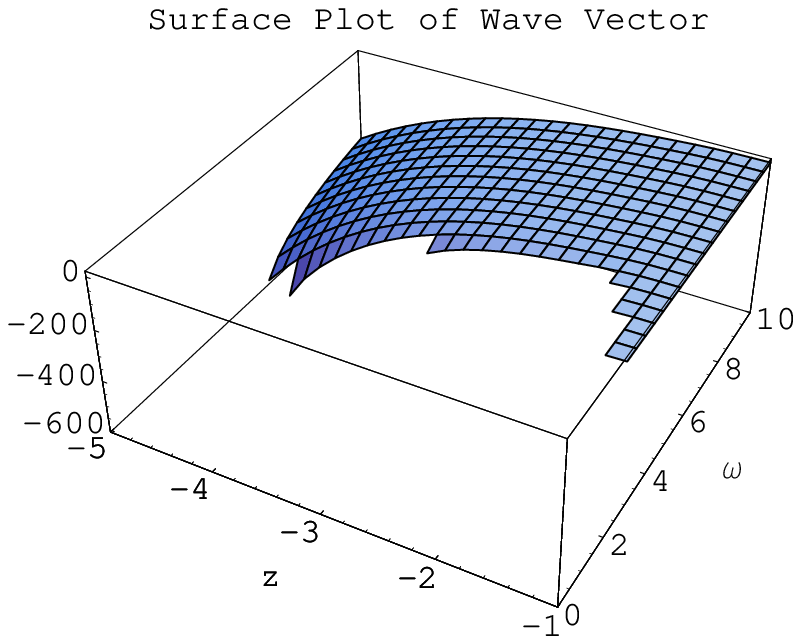,width=0.34\linewidth}
\epsfig{file=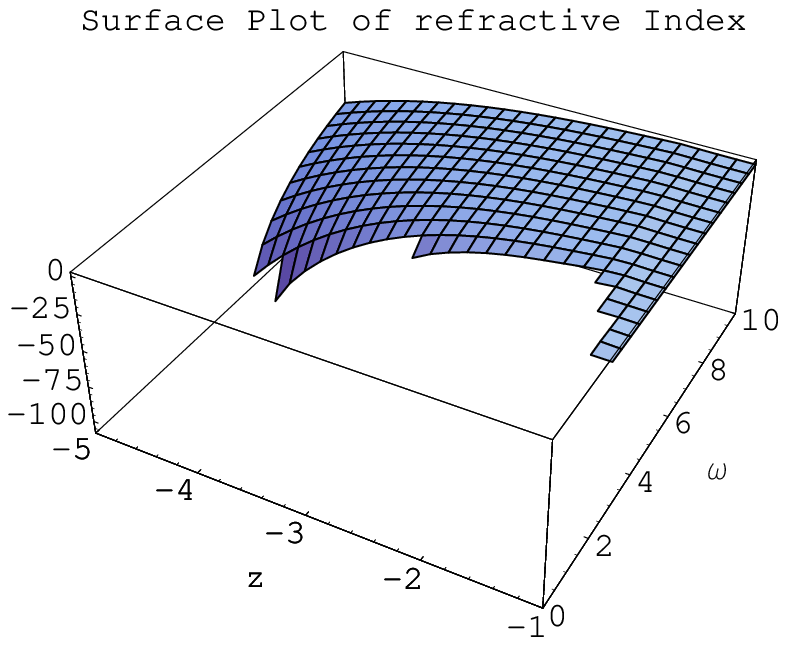,width=0.34\linewidth}\\
\epsfig{file=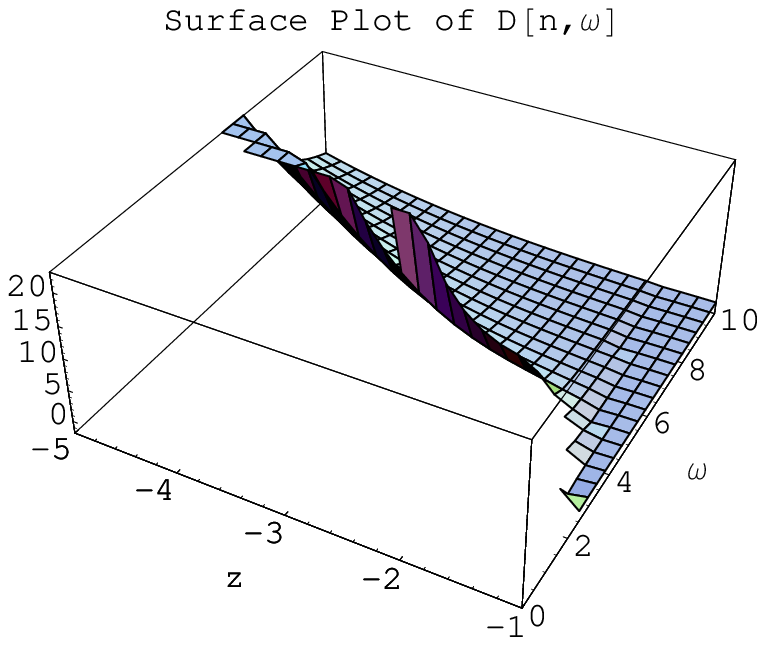,width=0.34\linewidth}
\epsfig{file=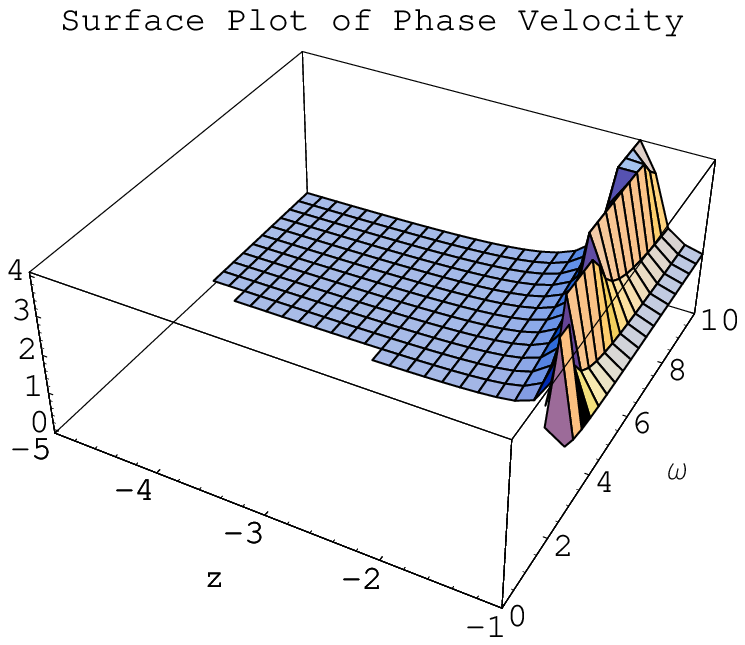,width=0.34\linewidth}
\epsfig{file=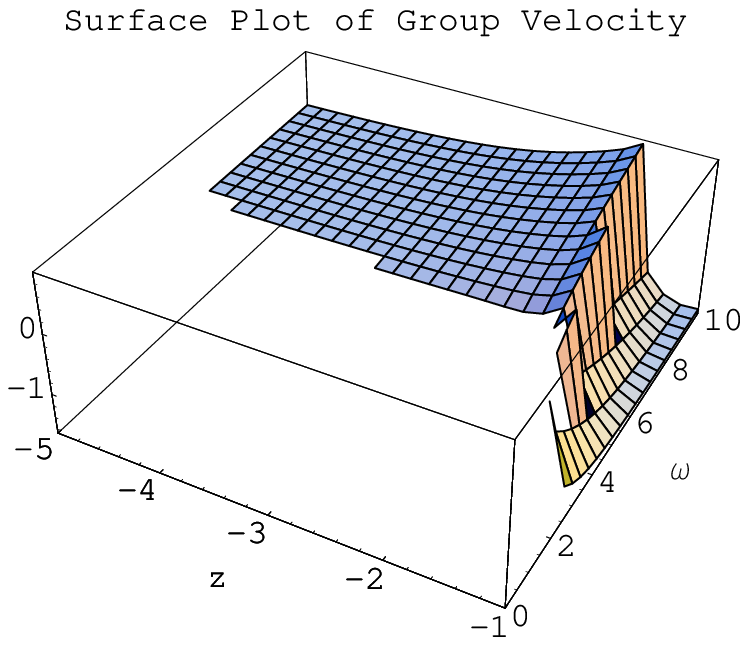,width=0.34\linewidth}\\
\end{tabular}
\caption{Whole region admits normal dispersion.}
\begin{tabular}{cc}\\
\epsfig{file=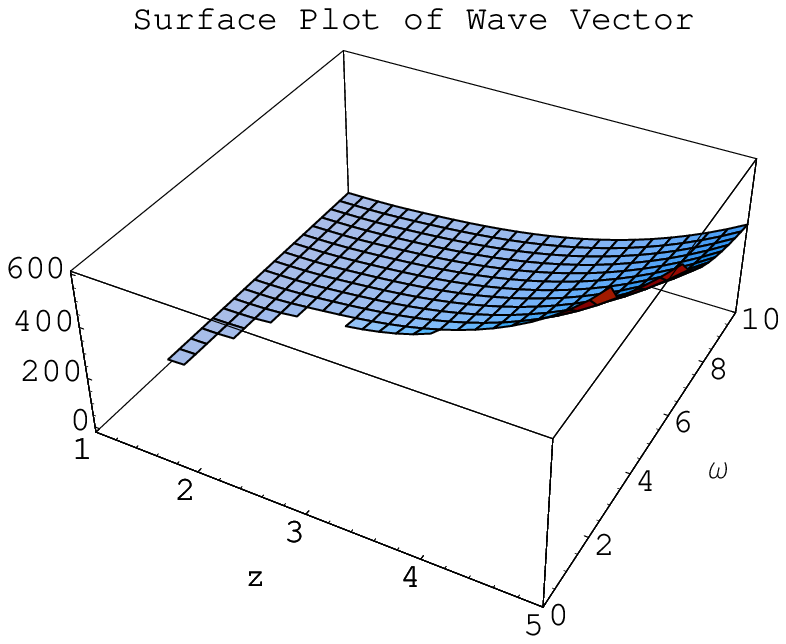,width=0.34\linewidth}
\epsfig{file=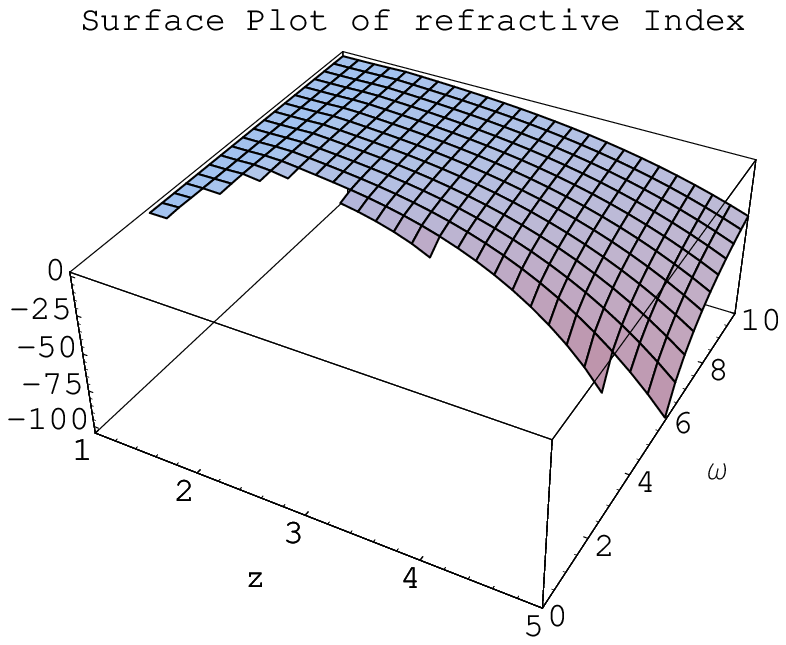,width=0.34\linewidth}\\
\epsfig{file=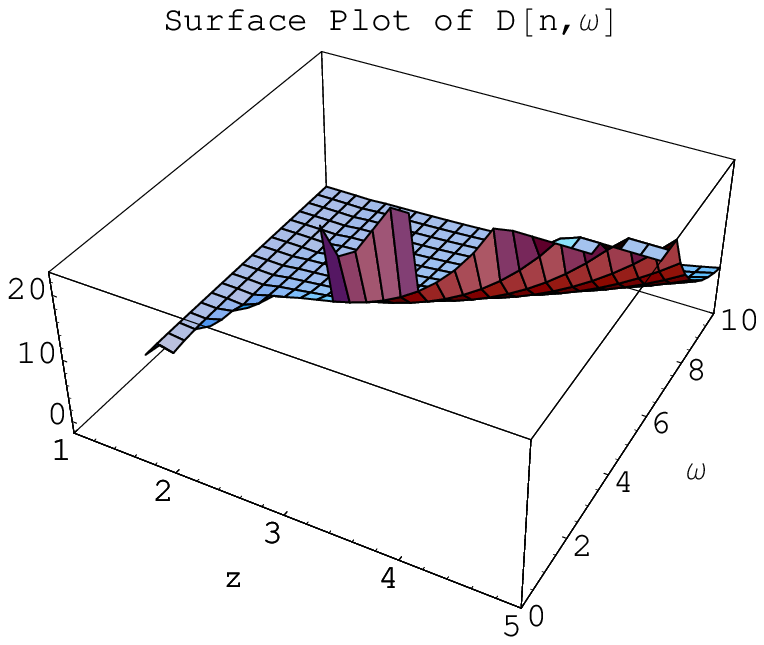,width=0.34\linewidth}
\epsfig{file=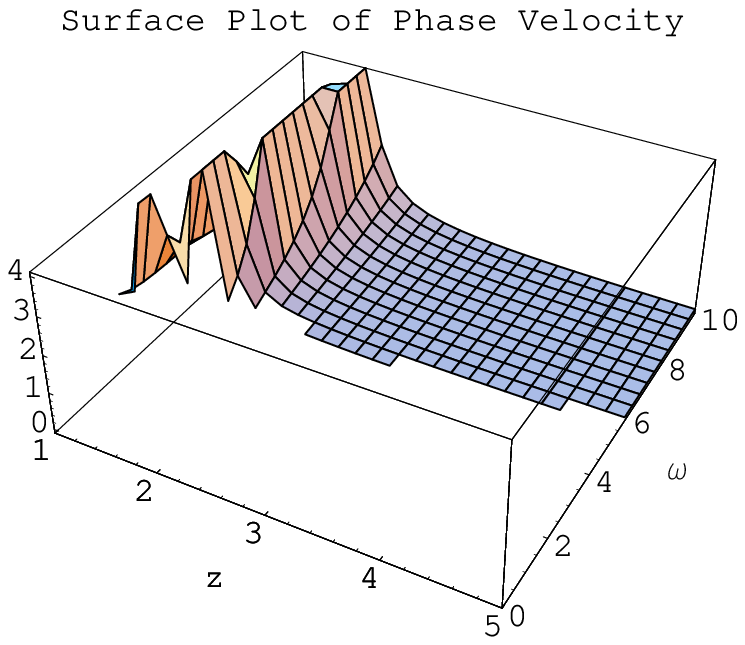,width=0.34\linewidth}
\epsfig{file=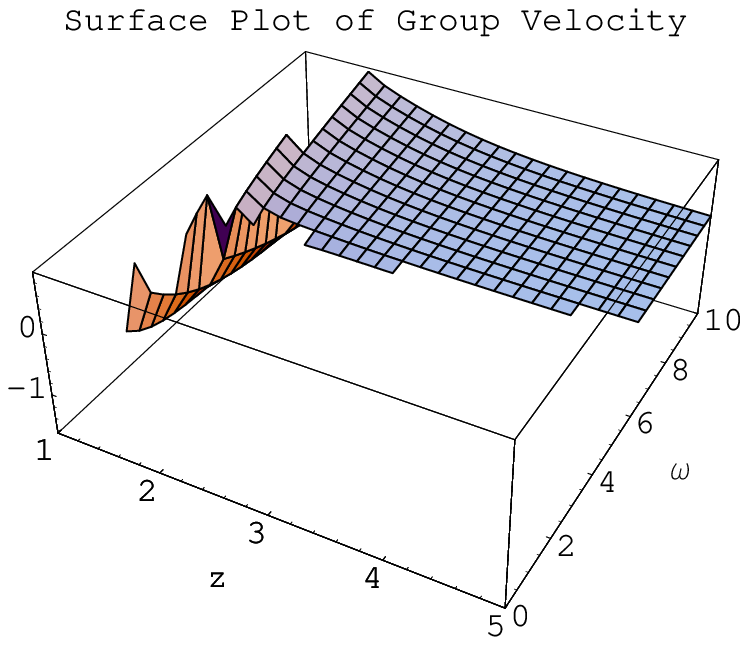,width=0.34\linewidth}\\
\end{tabular}
\caption{Dispersion of waves is normal throughout the region.}
\end{figure}
\begin{figure}
\begin{tabular}{cc}
\epsfig{file=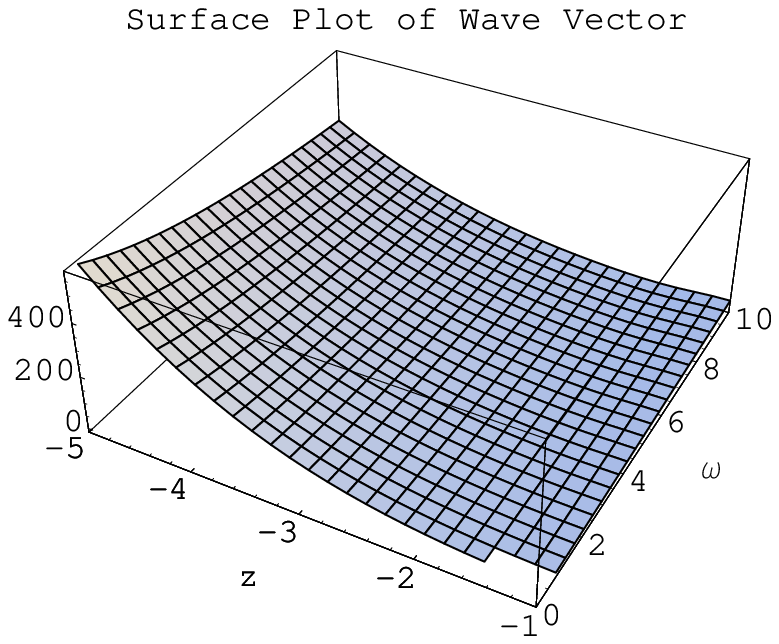,width=0.34\linewidth}
\epsfig{file=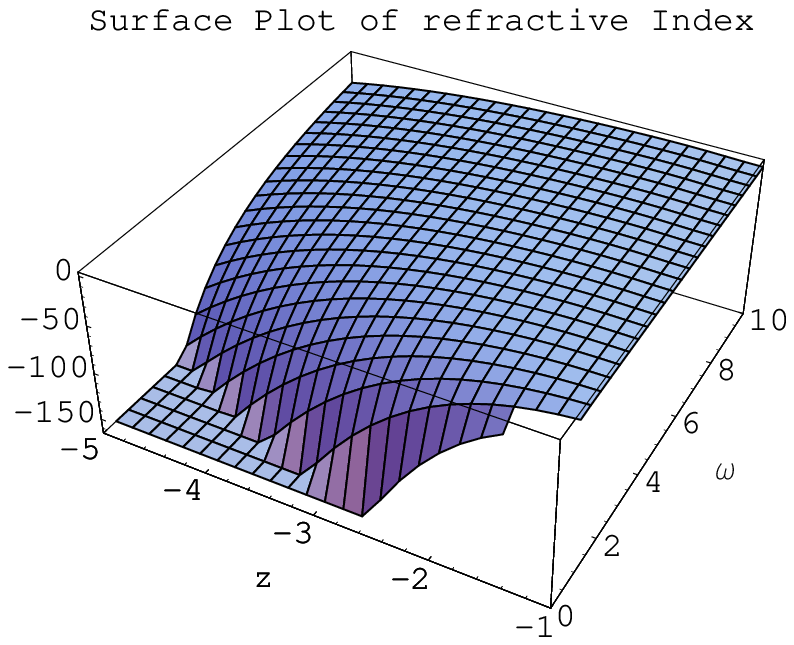,width=0.34\linewidth}\\
\epsfig{file=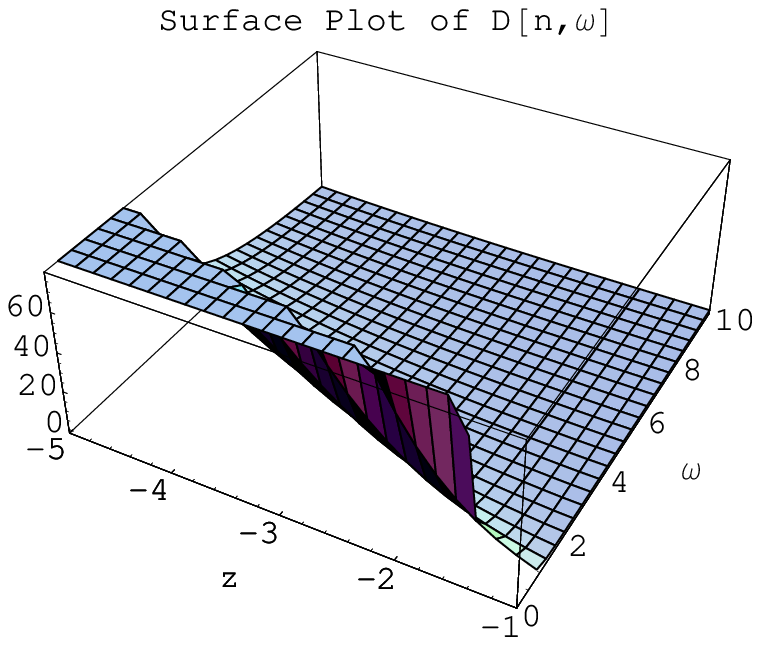,width=0.34\linewidth}
\epsfig{file=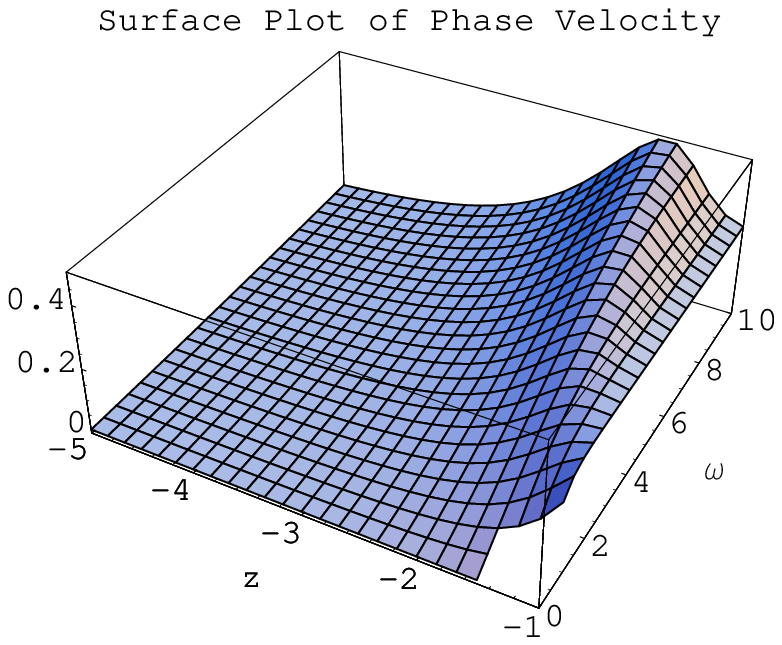,width=0.34\linewidth}
\epsfig{file=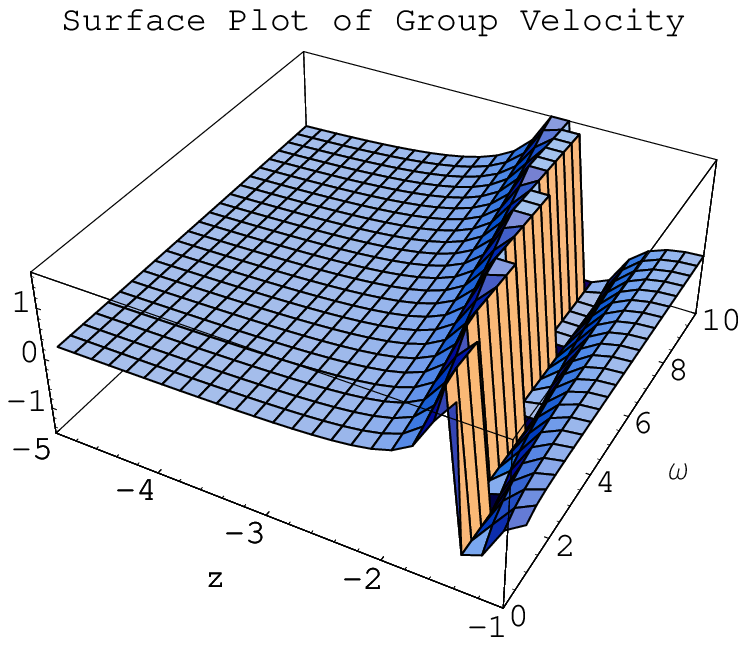,width=0.34\linewidth}\\
\end{tabular}
\caption{Waves disperse normally in the whole region.}
\begin{tabular}{cc}\\
\epsfig{file=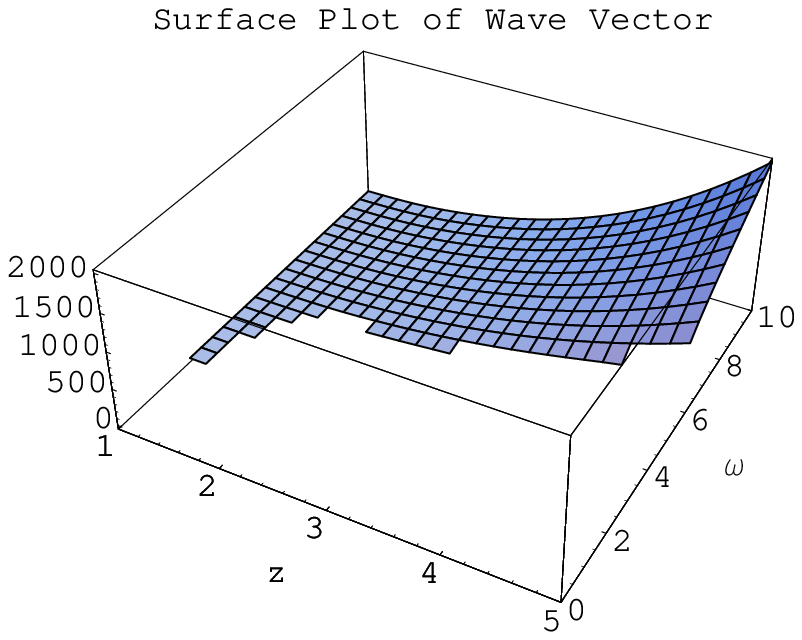,width=0.34\linewidth}
\epsfig{file=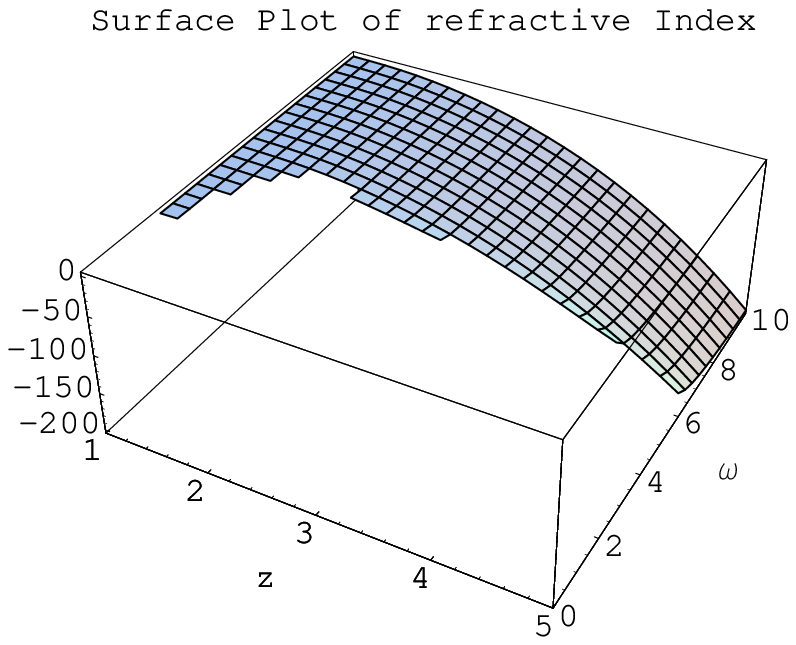,width=0.34\linewidth}\\
\epsfig{file=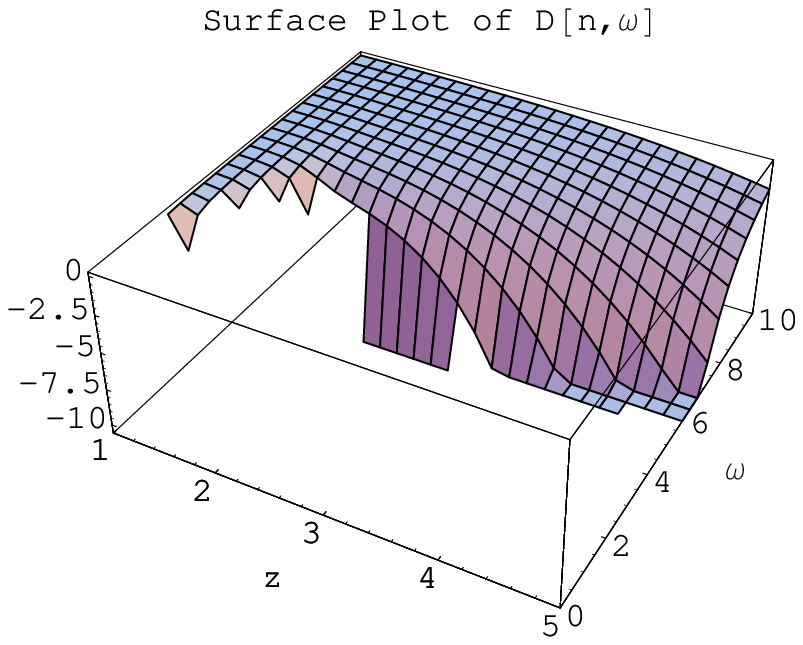,width=0.34\linewidth}
\epsfig{file=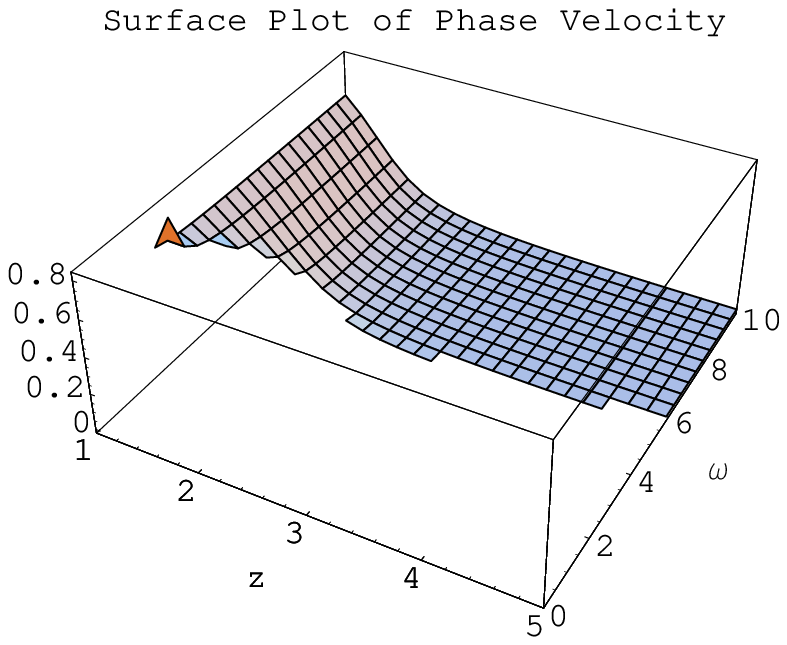,width=0.34\linewidth}
\epsfig{file=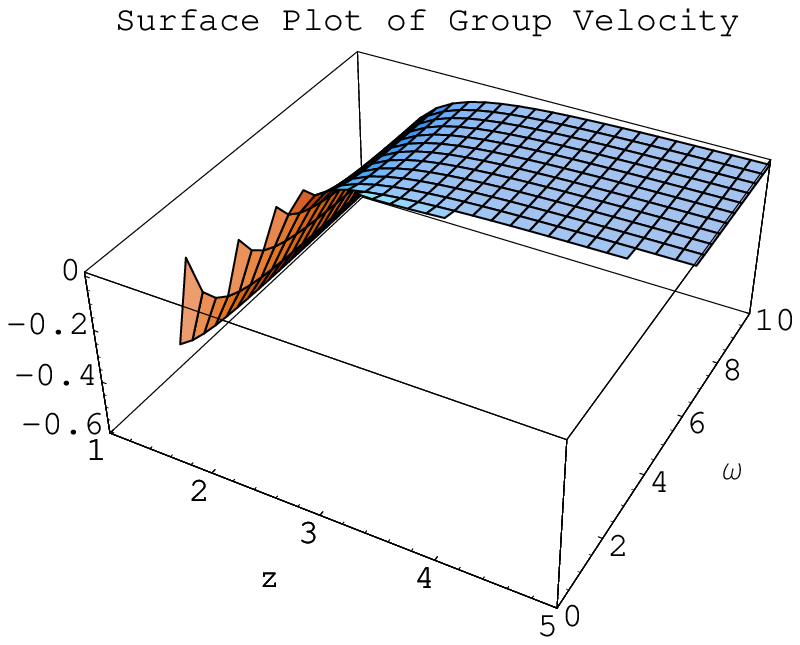,width=0.34\linewidth}\\
\end{tabular}
\caption{Region shows anomalous dispersion.}
\end{figure}
\begin{figure}
\begin{tabular}{cc}
\epsfig{file=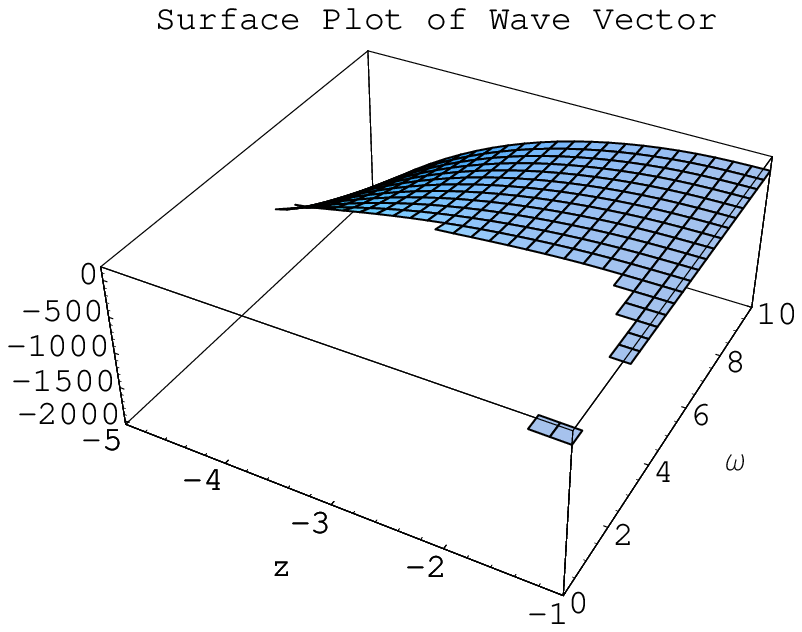,width=0.34\linewidth}
\epsfig{file=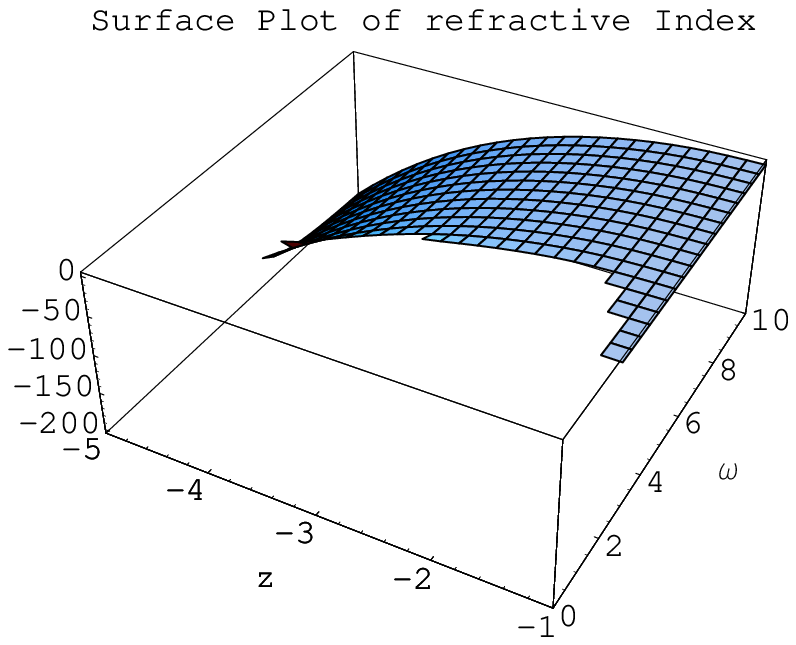,width=0.34\linewidth}\\
\epsfig{file=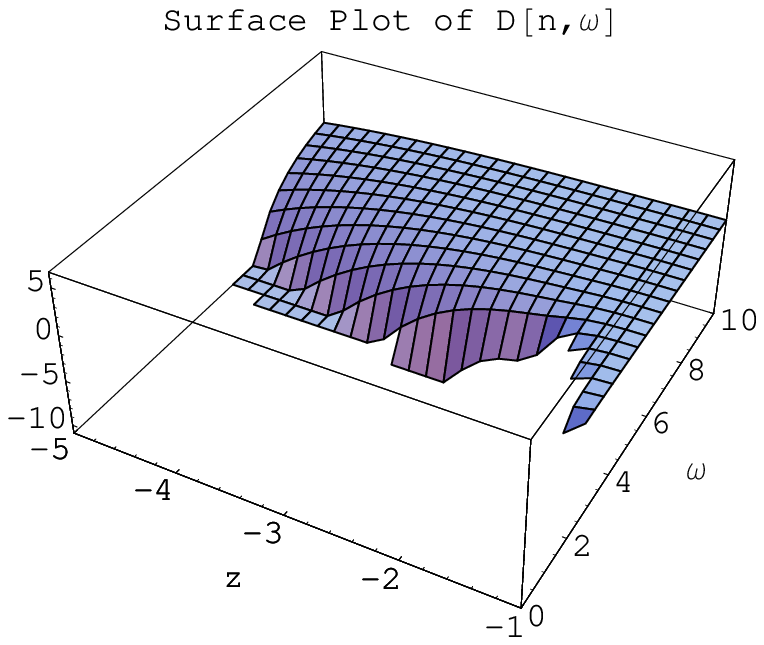,width=0.34\linewidth}
\epsfig{file=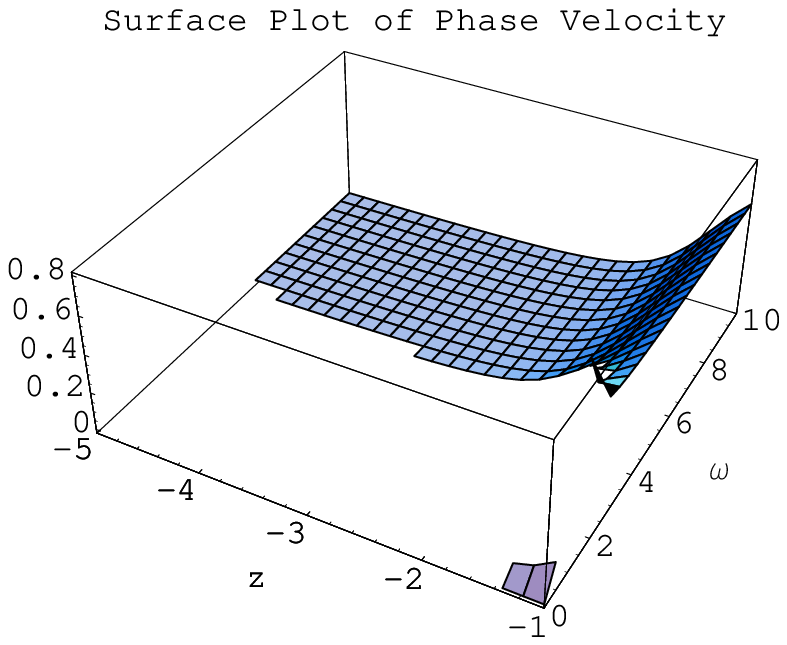,width=0.34\linewidth}
\epsfig{file=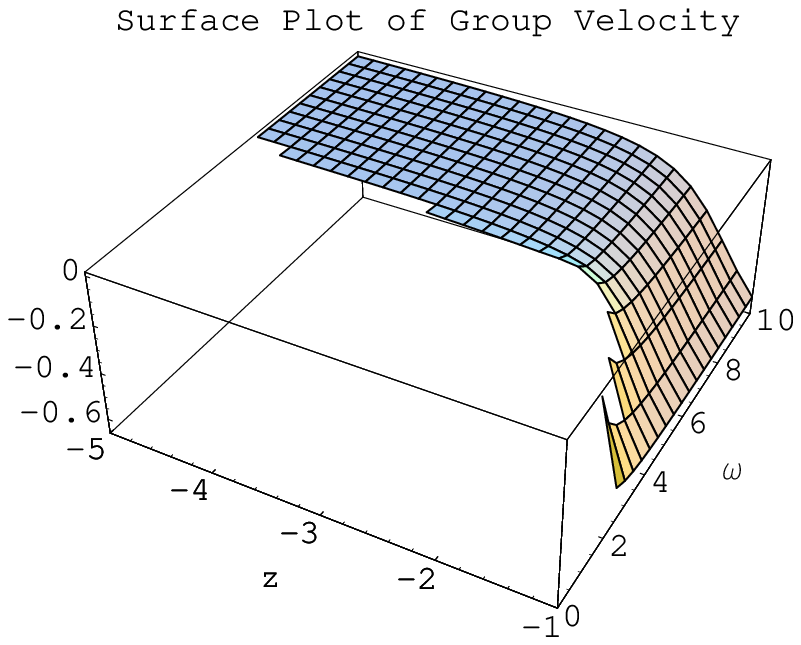,width=0.34\linewidth}\\
\end{tabular}
\caption{Waves exhibit both normal and anomalous dispersion.}
\begin{tabular}{cc}\\
\epsfig{file=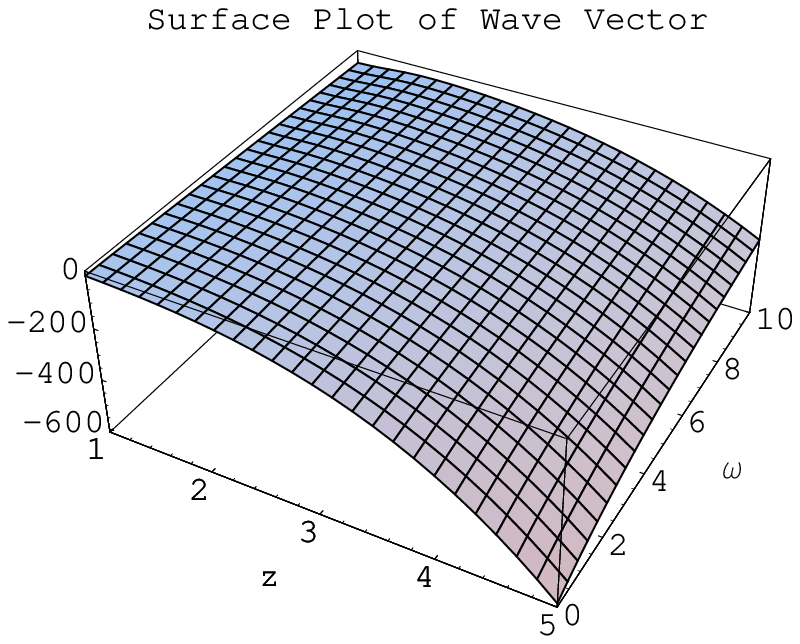,width=0.34\linewidth}
\epsfig{file=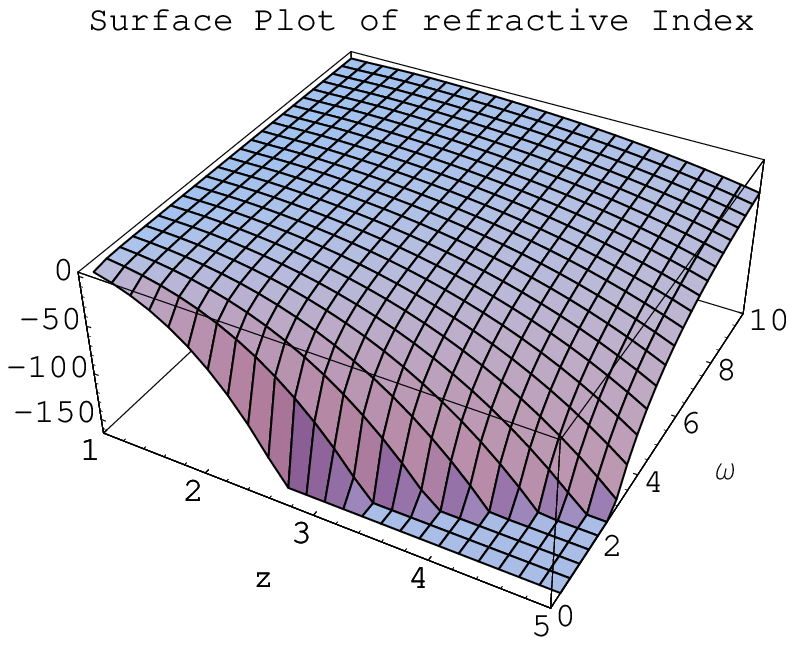,width=0.34\linewidth}\\
\epsfig{file=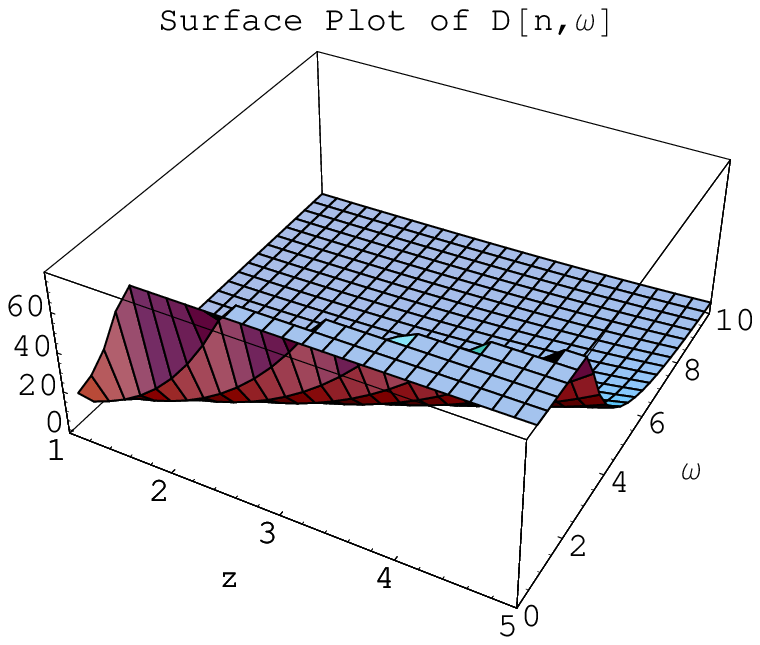,width=0.34\linewidth}
\epsfig{file=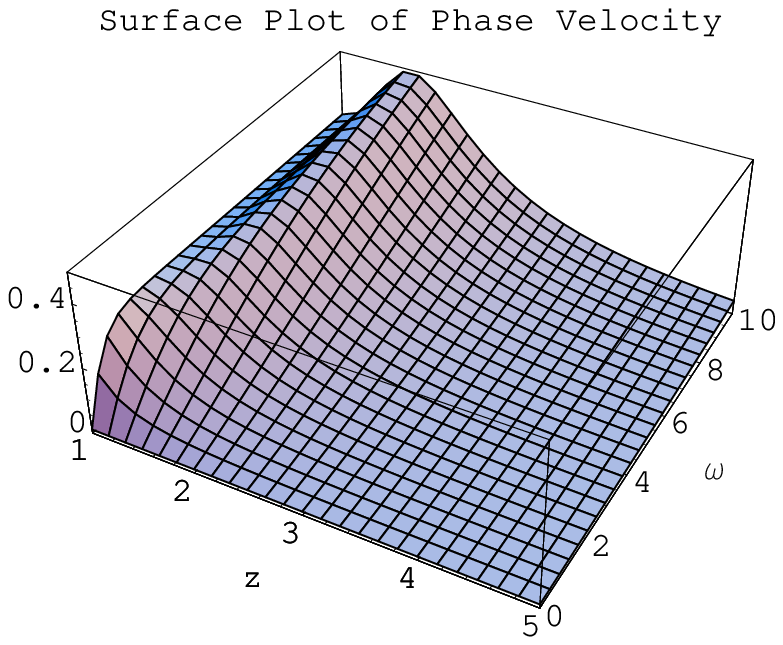,width=0.34\linewidth}
\epsfig{file=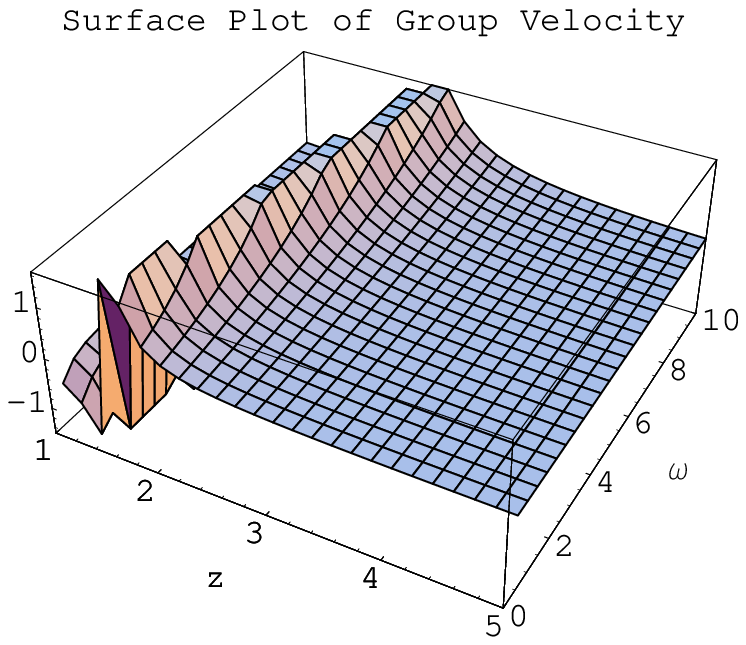,width=0.34\linewidth}\\
\end{tabular}
\caption{Dispersion is found to be normal in the whole region.}
\end{figure}
\begin{center}
\newpage
\par\noindent
Table II. Regions of dispersion
\end{center}
\begin{center}
\begin{tabular}{|c|c|c|c|c|}
\hline \textbf{Fig.}&  \textbf{ Normal dispersion} &
\textbf{Anomalous dispersion}
\\\hline  & $-5\leq z\leq -1.3, 0\leq\omega\leq 10$   &
$-1.28\leq z\leq -1.2, 3\leq\omega\leq 5$    \\
\textbf{1} & $-1.2\leq z\leq -1, 0\leq\omega\leq 1$   & $-1.28\leq
z\leq -1.2, 7\leq\omega\leq 10$
\\\hline & $-5\leq z\leq -4, 0\leq\omega\leq 10$       &
$-3\leq z\leq -2.5, 7\leq\omega\leq 8.5$     \\
\textbf{2} & $-4\leq z\leq -3, 0\leq\omega\leq 1$  & $-2.5\leq z\leq
-2, 7\leq\omega\leq 8.4$     \\& $-1.5\leq z\leq -1, 0\leq\omega\leq
1$   & $-2\leq z\leq -1.5, 2.5\leq\omega\leq 3$
\\\hline \textbf{3}& $3.9\leq z\leq 5, 0\leq\omega\leq 10$      &
$1.2\leq z\leq 3.6, 0\leq\omega\leq 10$
\\\hline & $1.5\leq z\leq
5, 0\leq\omega\leq 10$      &
$1\leq z\leq 1.1, 3\leq\omega\leq 4$ \\
\textbf{4} & $1\leq z\leq 1.5, 0\leq\omega\leq 2$   & $1.1\leq
z\leq 1.3, 5\leq\omega\leq 7$
\\\hline & $-5\leq z\leq -1, 0\leq\omega\leq 4.5$       &
$-5\leq z\leq -4, 8.2\leq\omega\leq 10$     \\
\textbf{9} & $-4\leq z\leq -3, 4.7\leq\omega\leq 5$  & $-4\leq
z\leq -3, 8.5\leq\omega\leq 10$
\\\hline
\end{tabular}
\end{center}

\section{Plasma Flow With Rotating Magnetized Background}

Here plasma is supposed to be rotating and magnetized. The
magnetic field and velocity of fluid are assumed to lie in
$xz$-plane. The corresponding perturbed Fourier analyzed GRMHD
equations, i.e., Eqs.(\ref{25})-(\ref{31}) are given in Section
$2$.

\subsection{Numerical Solutions}

We take the same assumptions for the lapse function, velocity and
specific enthalpy as in the previous section. Further, we assume
$\frac{B^{2}}{4\pi}=2$ with $u=V$ and $V^F=1$ in Eq.(\ref{a}) so
that $\lambda=1+\frac{\sqrt{2+z^{2}}}{z}$.

Here we also consider the region $-5\leq z\leq5,~ 0\leq\omega\leq
10$ and investigate the wave properties in meshes $-5\leq z\leq-1$
and $1\leq z\leq5$ From Eqs.(\ref{26})-(\ref{27}), it follows that
$c_{5}=0$. Consequently, we obtain dispersion relation whose real
part is
\begin{equation}{\setcounter{equation}{1}}\label{38}
A_1(z)k^4+A_2(z,\omega)k^3+A_3(z,\omega)k^2+A_4(z,\omega)k+A_5(z,\omega)=0
\end{equation}
giving four imaginary roots. The imaginary part of the dispersion
relation
\begin{eqnarray}\label{39}
&&B_1(z)k^5+B_2(z,\omega)k^4+B_3(z,\omega)k^3+B_4(z,\omega)k^2+B_5(z,\omega)k\nonumber\\
&&+B_6(z,\omega)=0
\end{eqnarray}
yields five roots of $k$ out of which one is real and four roots
are complex. The real root indicates wave propagation in both
meshes, i.e., $-5\leq z\leq-1$ and $1\leq z\leq5$ shown in Figures
\textbf{11}-\textbf{12}. This shows that waves move towards the
event horizon. Also, it is obvious from figures that dispersion is
normal as well as anomalous at random points.

The following tables show the results obtained from these figures.
\begin{center}
Table III. Direction and refractive index of waves
\end{center}
\begin{tabular}{|c|c|c|c|c|}
\hline\textbf{Fig.} & \textbf{Direction of Waves} &
\textbf{Refractive Index} ($n$)\\ \hline
& & $n<1$ and decreases in the region \\
\textbf{11} & Move towards the event horizon & $-5\leq z\leq
-2.1,0\leq\omega\leq 10$\\&& with the decrease in $z$  \\
\hline
& & $n<1$ and increases in the region\\
\textbf{12} & Move towards the event horizon & $1\leq z\leq 1.8,
1.5\leq\omega\leq
4$\\&&with the decrease in $z$  \\
\hline
\end{tabular}\\
\begin{center}
\begin{figure}
\begin{tabular}{cc}
\epsfig{file=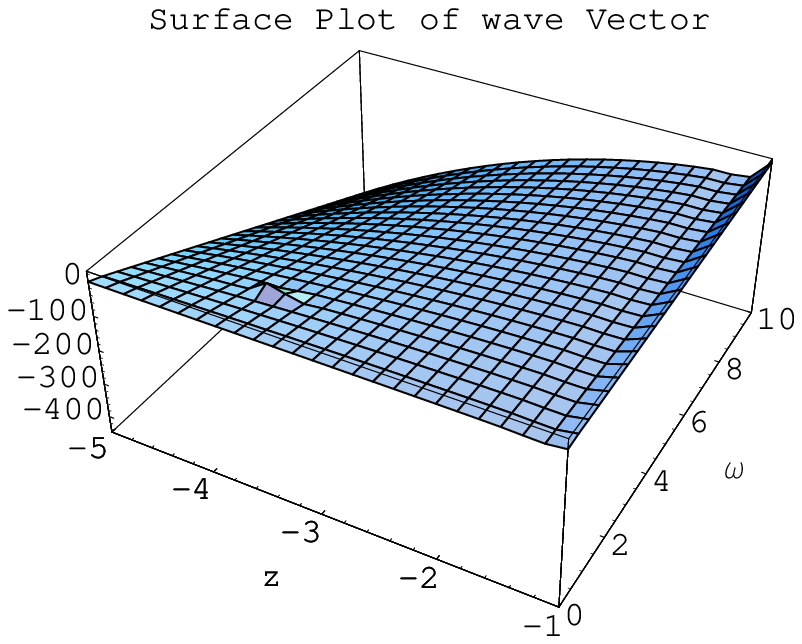,width=0.34\linewidth}
\epsfig{file=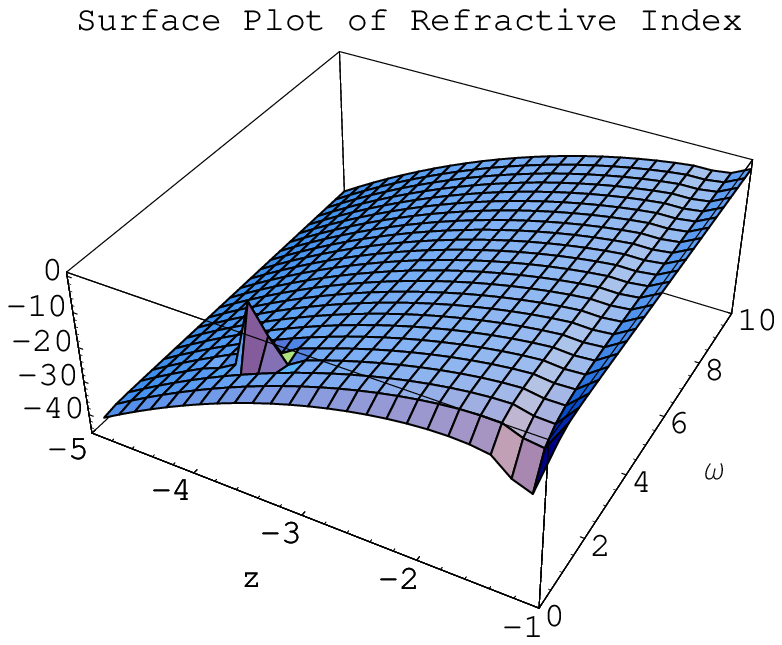,width=0.34\linewidth}\\
\epsfig{file=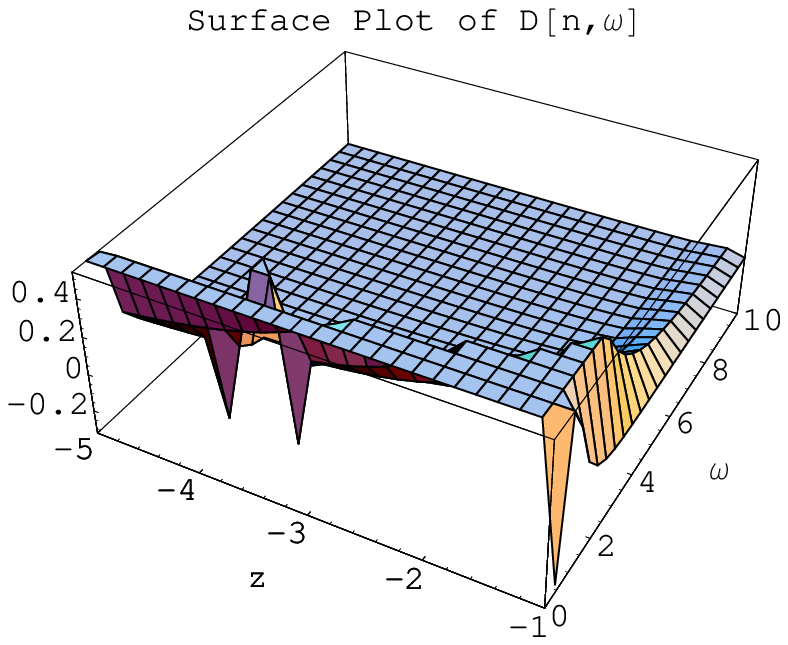,width=0.34\linewidth}
\epsfig{file=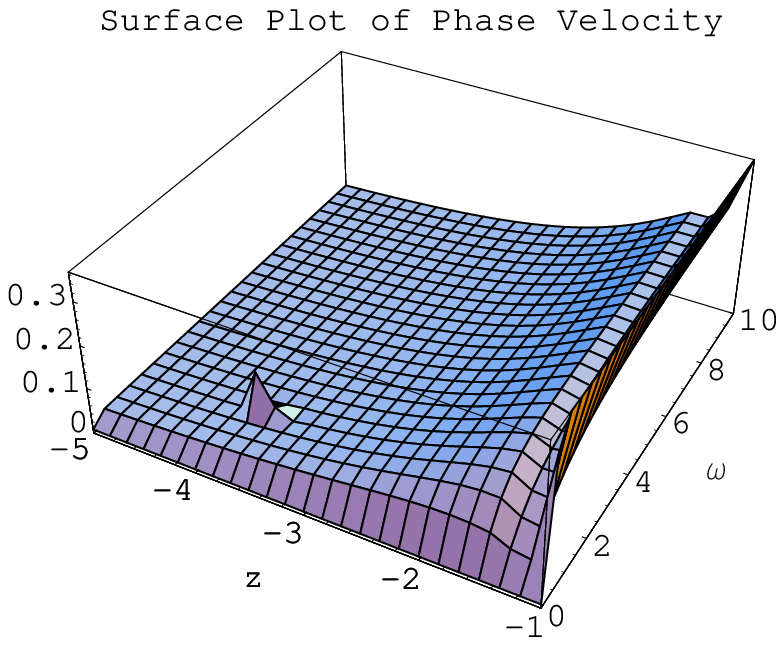,width=0.34\linewidth}
\epsfig{file=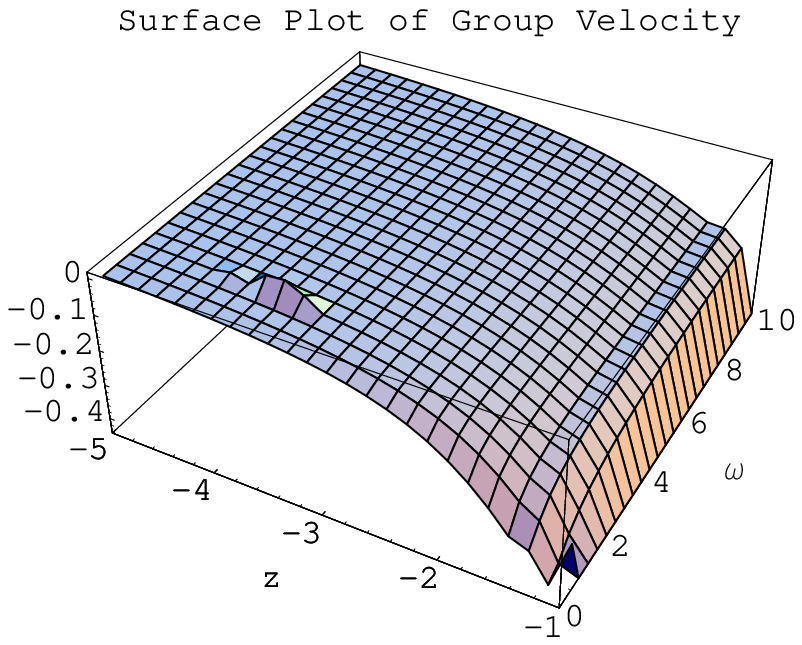,width=0.34\linewidth}\\
\end{tabular}
\caption{Normal and anomalous dispersion at random points.}
\begin{tabular}{cc}\\
\epsfig{file=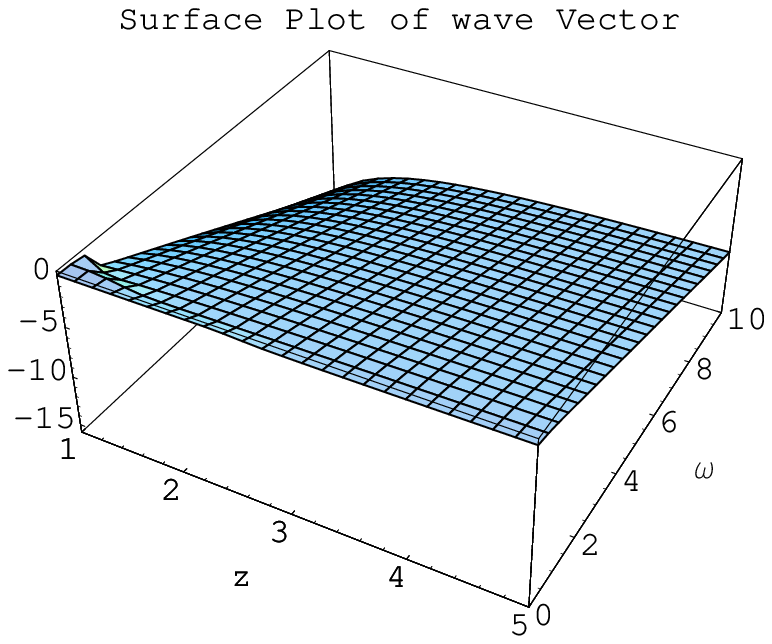,width=0.34\linewidth}
\epsfig{file=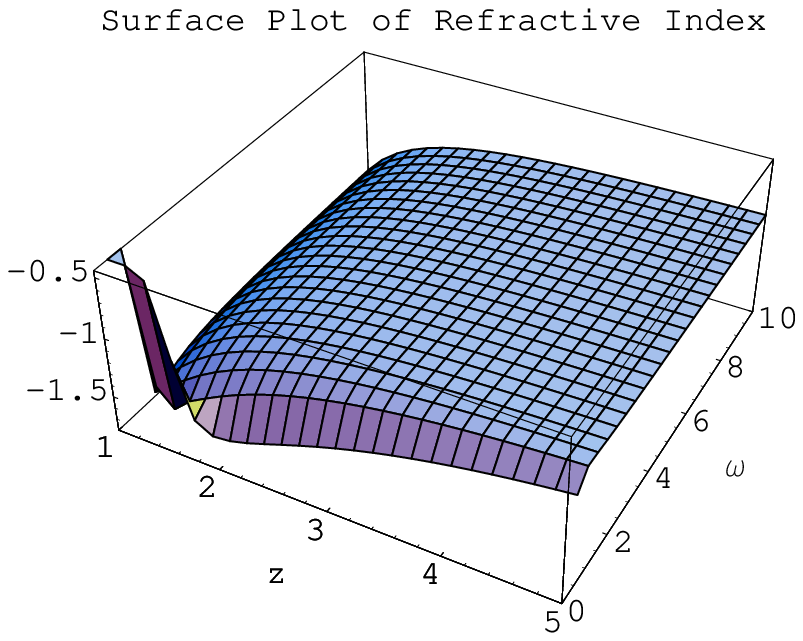,width=0.34\linewidth}\\
\epsfig{file=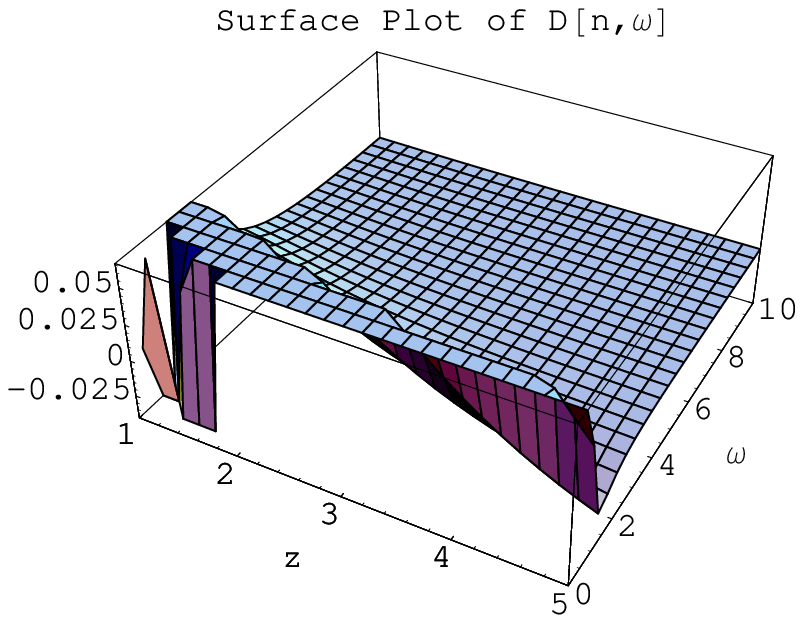,width=0.34\linewidth}
\epsfig{file=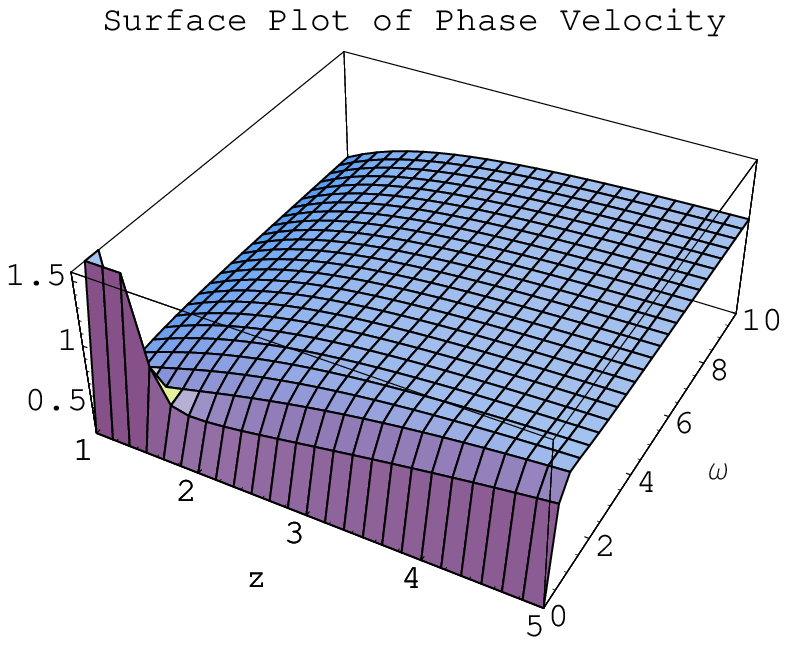,width=0.34\linewidth}
\epsfig{file=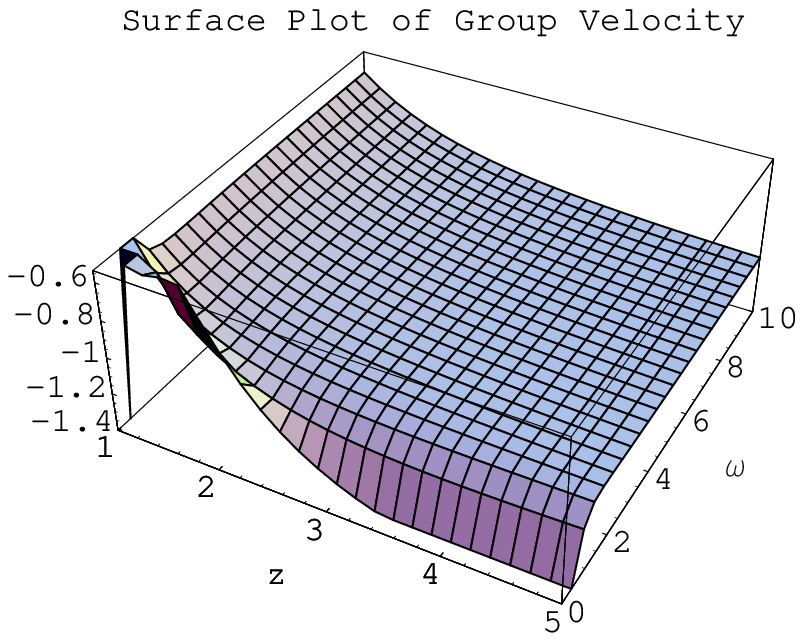,width=0.34\linewidth}\\
\end{tabular}
\caption{Dispersion is normal as well as anomalous at random
points.}
\end{figure}

Table IV. Regions of dispersion
\end{center}
\begin{center}
\begin{tabular}{|c|c|c|c|c|}
\hline \textbf{Fig.} & \textbf{ Normal dispersion} &
\textbf{Anomalous dispersion}
\\\hline & $-5\leq z\leq -4.9, 1\leq\omega\leq 10$      &
$-4\leq z\leq -3.6, 1\leq\omega\leq 1.5$ \\
\textbf{11} & $-4\leq z\leq -3, 2.5\leq\omega\leq 10$   &
$-3.7\leq z\leq -3.5, 1.9\leq\omega\leq 2.1$    \\
& $-3\leq z\leq -1, 2.8\leq\omega\leq 10$  & $-3.5\leq z\leq
-3.35, 1.8\leq\omega\leq 2.2$
\\\hline
& $1\leq z\leq 2, 2\leq\omega\leq 10$  &
$1\leq z\leq 2, 0.8\leq\omega\leq 1.1$     \\
\textbf{12}& $2\leq z\leq 4.5, 5\leq\omega\leq 10$  &
$4\leq z\leq 4.5, 1.4\leq\omega\leq 4$     \\
& $4.1\leq z\leq 4.5, 5.1\leq\omega\leq 10$   & $4\leq z\leq 4.5,
1.4\leq\omega\leq 4.5$
\\\hline
\end{tabular}
\end{center}

\section{Summary}

This paper deals with the study of isothermal plasma wave properties
in magnetosphere of SdS black hole in a Veselago medium. The ADM
$3+1$ formalism has been used to formulate the GRMHD equations for
this unusual medium. We have applied linear perturbations to the
GRMHD equations and have obtained their component form with the
assumption that plasma flows in two dimensions. Finally, we have
obtained dispersion relations for the rotating (non-magnetized and
magnetized) background.

For the rotating non-magnetized background, waves move towards the
event horizon shown in Figures \textbf{1}, \textbf{3}, \textbf{5},
\textbf{9} and \textbf{10} while waves are directed away from the
event horizon in Figures \textbf{2}, \textbf{4}, \textbf{6},
\textbf{7} and \textbf{8}. The dispersion is found to be normal as
well as anomalous at random points in Figures \textbf{1},
\textbf{2}, \textbf{3}, \textbf{4} and \textbf{9}. The Figures
\textbf{5}, \textbf{6}, \textbf{7} and \textbf{10} show normal
dispersion while \textbf{8} admits anomalous dispersion in the whole
region. The Figures \textbf{11} and \textbf{12} indicate that waves
are directed towards the event horizon for rotating magnetized
plasma. It is clear from these figures that region admits normal and
anomalous dispersion at random points.

We know that the refractive index is always greater than one in
the usual medium, while it is less than one for the Veselago
medium. Here we have found that the refractive index is less than
one and increases in small regions. The phase velocity is greater
than group velocity for both non-magnetized and magnetized
backgrounds. These are prominent aspects of the Veselago medium
which confirms the presence of this unusual medium for both
rotating (non-magnetized and magnetized) plasma in SdS black hole.

It is interesting to mention here that in a recent work (Sharif
and Mukthar 2011a, 2011b) for isothermal plasma on Schwarzschild
black hole, there does not exist waves for the rotating magnetized
plasma. However, we have seen wave propagation in SdS black hole
for this case. Here waves admit normal dispersion at most of
points while for the schwarzschild black hole, most of the waves
disperse anomalously. Thus it can be concluded that more
information can be extracted from magnetosphere by inclusion of
the de-Sitter patch in the Schwarzschild spacetime. It would be
interesting to extend this analysis for hot plasma which is in
progress.

\renewcommand{\theequation}{A\arabic{equation}}

\section*{Appendix}

The Maxwell equations, the $3+1$ GRMHD equations for the SdS
spacetime are given in this appendix. The Maxwell equations for such
a medium are
\begin{eqnarray}{\setcounter{equation}{1}}
\label{40}&&\nabla.\textbf{B}=0,\\
\label{41}&&\nabla\times\textbf{E}+\frac{\partial\textbf{B}}{\partial
t}=0,\\
\label{42}&&\nabla\cdot\textbf{E}=-\frac{\rho_e}{\epsilon},\\
\label{43}&&\nabla\times\textbf{B}=-\mu\textbf{j}+\frac{\partial\textbf{E}}{\partial
t}=0.
\end{eqnarray}

The GRMHD equations for the SdS spacetime in Rindler coordinates
turn out to be (Sharif and Mukthar 2011a, 2011b)
\begin{eqnarray}\label{49}
&&\frac{\partial\textbf{B}}{\partial t}=-\nabla \times(\alpha
\textbf{V}\times\textbf{B}),\\\label{50}
&&\nabla.\textbf{B}=0,\\\label{51}
&&\frac{\partial\rho_0}{\partial
t}+(\alpha\textbf{V}.\nabla)\rho_0+\rho_0\gamma^2
\textbf{V}.\frac{\partial\textbf{V}}{\partial
t}+\rho_0\gamma^2\textbf{V}.(\alpha\textbf{V}.\nabla)\textbf{V}\nonumber\\
&&+\rho_0{\nabla.(\alpha\textbf{V})}=0, \\\label{52}
&&\{(\rho_0\mu\gamma^2+\frac{\textbf{B}^2}{4\pi})\delta_{ij}
+\rho_0\mu\gamma^4V_iV_j
-\frac{1}{4\pi}B_iB_j\}(\frac{1}{\alpha}\frac{\partial}{\partial
t}+\textbf{V}.\nabla)V^j\nonumber\\
&&-(\frac{\textbf{B}^2}{4\pi}\delta_{ij}-\frac{1}{4\pi}B_iB_j)
V^j,_kV^k+\rho_0\gamma^2V_i\{\frac{1}{\alpha}\frac{\partial
\mu}{\partial t}+(\textbf{V}.\nabla)\mu\}\nonumber\\
&&=-\rho_0\mu\gamma^2a_i-p,_i+
\frac{1}{4\pi}(\textbf{V}\times\textbf{B})_i\nabla.(\textbf{V}\times\textbf{B})
-\frac{1}{8\pi\alpha^2}(\alpha\textbf{B})^2,_i\nonumber\\
&&+\frac{1}{4\pi\alpha}(\alpha B_i),_jB^j-\frac{1}{4\pi\alpha}
[\textbf{B}\times\{\textbf{V}\times(\nabla\times(\alpha\textbf{V}
\times\textbf{B}))\}]_i,
\\\label{53}
&&(\frac{1}{\alpha}\frac{\partial}{\partial
t}+\textbf{V}.\nabla)(\mu\rho_0\gamma^2)-\frac{1}{\alpha}\frac{\partial
p }{\partial
t}+2\mu\rho_0\gamma^2(\textbf{V}.\textbf{a})+\mu\rho_0\gamma^2
(\nabla.\textbf{V})\nonumber\\&&-\frac{1}{4\pi}
(\textbf{V}\times\textbf{B}).(\textbf{V}\times\frac{1}{\alpha}\frac{\partial
\textbf{B}}{\partial t})-\frac{1}{4\pi}
(\textbf{V}\times\textbf{B}).(\textbf{B}\times\frac{1}{\alpha}\frac{\partial
\textbf{V}}{\partial
t})\nonumber\\&&+\frac{1}{4\pi\alpha}\left(\textbf{V}\times\textbf{B}).
(\nabla\times\alpha\textbf{B}\right.)=0.
\end{eqnarray}

\end{document}